\documentclass[10pt,twoside,journal,compsoc]{IEEEtran}
\ifCLASSOPTIONcompsoc
  \usepackage[nocompress,noadjust]{cite}
\else
  \usepackage{cite}
\fi
\ifCLASSINFOpdf
  \usepackage[pdftex]{graphicx}
\else
  \usepackage[dvips]{graphicx}
\fi
\pdfpageattr{/Group << /S /Transparency /I true /CS /DeviceRGB>>}

\usepackage{mathptmx} 
\usepackage{amsmath}
\usepackage{amssymb}  

\catcode`\_=12
{\catcode`\_=\active \gdef_#1{\sb{\mathrm{#1}}}}

\usepackage[euler]{textgreek}  
\usepackage{siunitx}
\usepackage{color}

\usepackage{multirow}           
\usepackage{booktabs}
\usepackage{tabularx}
\newcolumntype{L}[1]{>{\raggedright\arraybackslash}p{#1}} 
\newcolumntype{C}[1]{>{\centering\arraybackslash}p{#1}}   
\newcolumntype{R}[1]{>{\raggedleft\arraybackslash}p{#1}}  
\newcommand{\ctab}{\centering\arraybackslash} 

\usepackage{xspace}
\usepackage{paralist}

\usepackage[toc, section=section]{glossaries}  
\makeatletter
\newcommand{\myacronymentry}[5]{
\newglossaryentry{#1}{
  type = #2,
  name = #3,
  description = #4,
  first = #4 (#3),
  text = #3,
  plural = #3\glspluralsuffix,
  firstplural = #4\glspluralsuffix\space (#3\glspluralsuffix),
  sort = #5}
}

\makeatother
\newglossary[glsAcrLog]{acronyms}{glsAcrIn}{glsAcrOut}{acronyms}

\makeglossaries

\myacronymentry{AES}{\glsdefaulttype}{AES}{Advanced Encryption Standard}{aes}
\myacronymentry{ASIC}{\glsdefaulttype}{ASIC}{application-specific integrated circuit}{asic}

\myacronymentry{CMOS}{\glsdefaulttype}{CMOS}{complementary metal-oxide semiconductor}{cmos}
\myacronymentry{CDMA}{\glsdefaulttype}{CDMA}{cluster DMA}{cdma}
\myacronymentry{CMP}{\glsdefaulttype}{CMP}{chip multiprocessor}{cmp}

\myacronymentry{DAC}{\glsdefaulttype}{DAC}{digital-to-analog converter}{dac}
\myacronymentry{DMA}{\glsdefaulttype}{DMA}{direct memory access unit}{dma}
\myacronymentry{DSP}{\glsdefaulttype}{DSP}{digital signal processing}{dsp}
\myacronymentry{DWT}{\glsdefaulttype}{DWT}{discrete wavelet transform}{dwt}

\myacronymentry{FANN}{\glsdefaulttype}{FANN}{fast artificial neural network}{fann}
\myacronymentry{FFT}{\glsdefaulttype}{FFT}{fast Fourier transform}{fft}
\myacronymentry{FIFO}{\glsdefaulttype}{FIFO}{first-in first-out}{fifo}
\myacronymentry{FOM}{\glsdefaulttype}{FoM}{figure-of-merit}{fom}
\myacronymentry{FSM}{\glsdefaulttype}{FSM}{finite state machine}{fsm}
\myacronymentry{FPGA}{\glsdefaulttype}{FPGA}{field programmable gate array}{fpga}

\myacronymentry{GE}{\glsdefaulttype}{GE}{gate equivalent}{ge}
\myacronymentry{GPU}{\glsdefaulttype}{GPU}{graphics processing unit}{gpu}
\myacronymentry{GP-GPU}{\glsdefaulttype}{GP-GPU}{general-purpose GPU}{gl-gpu}

\myacronymentry{HDL}{\glsdefaulttype}{HDL}{hardware description language}{hdl}
\myacronymentry{HPF}{\glsdefaulttype}{HPF}{high-pass filter}{hpf}

\myacronymentry{IoT}{\glsdefaulttype}{IoT}{internet-of-things}{iot}
\myacronymentry{ISA}{\glsdefaulttype}{ISA}{instruction set architecture}{isa}

\myacronymentry{L1}{\glsdefaulttype}{L1}{level 1}{l1}
\myacronymentry{L2}{\glsdefaulttype}{L2}{level 2}{l2}
\myacronymentry{LINT}{\glsdefaulttype}{LINT}{logarithmic interconnect}{lint}
\myacronymentry{LSB}{\glsdefaulttype}{LSB}{least significant bit}{lsb}
\myacronymentry{LPF}{\glsdefaulttype}{LPF}{low-pass filter}{lpf}

\myacronymentry{MFC}{\glsdefaulttype}{MFC}{Mel-frequency cepstrum}{mfc}
\myacronymentry{MCU}{\glsdefaulttype}{MCU}{microcontroller unit}{mcu}
\myacronymentry{MOSFET}{\glsdefaulttype}{MOSFET}{metal-oxide-semiconductor field-effect transistor}{mosfet}
\myacronymentry{MMU}{\glsdefaulttype}{MMU}{memory management unit}{mmu}

\myacronymentry{NoC}{\glsdefaulttype}{NoC}{network-on-chip}{noc}
\myacronymentry{NTC}{\glsdefaulttype}{NTC}{near-threshold computing}{ntc}

\myacronymentry{OEP}{\glsdefaulttype}{OEP}{optimal energy point}{oep}

\myacronymentry{PCA}{\glsdefaulttype}{PCA}{principal component analysis}{pca}
\myacronymentry{PE}{\glsdefaulttype}{PE}{processing element}{pe}
\myacronymentry{PULP}{\glsdefaulttype}{PULP}{parallel ultra-low power}{pulp}
\myacronymentry{PM}{\glsdefaulttype}{PM}{power management}{pm}
\myacronymentry{PDMA}{\glsdefaulttype}{PDMA}{peripheral DMA}{pdma}
\myacronymentry{PMIC}{\glsdefaulttype}{PMIC}{power management IC}{pmic}
\myacronymentry{PMCA}{\glsdefaulttype}{PMCA}{programmable many-core accelerator}{pmca}

\myacronymentry{ROM}{\glsdefaulttype}{ROM}{read-only memory}{rom}
\myacronymentry{RMS}{\glsdefaulttype}{RMS}{root mean square}{rms}
\myacronymentry{Rx}{\glsdefaulttype}{Rx}{receiver}{rx}
\myacronymentry{RnD}{\glsdefaulttype}{R\&{}D}{research and development}{retd}
\myacronymentry{RTL}{\glsdefaulttype}{RTL}{register-level transfer}{rtl}
\myacronymentry{RF}{\glsdefaulttype}{RF}{radio frequency}{rf}

\myacronymentry{SB}{\glsdefaulttype}{SB}{synchronization-operation buffer}{sb}
\myacronymentry{SCM}{\glsdefaulttype}{SCM}{standard cell memory}{scm}
\myacronymentry{SCU}{\glsdefaulttype}{SCU}{synchronization and communication unit}{scm}
\myacronymentry{SFR}{\glsdefaulttype}{SFR}{synchronization-free region}{sfr}
\myacronymentry{SFDR}{\glsdefaulttype}{SFDR}{spurious free dynamic range}{sfdr}
\myacronymentry{SNR}{\glsdefaulttype}{SNR}{signal-to-noise ratio}{snr}
\myacronymentry{SDR}{\glsdefaulttype}{SDR}{signal-to-distortion ratio}{sdr}
\myacronymentry{SPI}{\glsdefaulttype}{SPI}{serial peripheral interface}{spi}
\myacronymentry{SPM}{\glsdefaulttype}{SPM}{scratchpad memory}{spm}
\myacronymentry{SoC}{\glsdefaulttype}{SoC}{system-on-chip}{soc}

\myacronymentry{SIMD}{\glsdefaulttype}{SIMD}{single-instruction\, multiple-data}{simd}
\myacronymentry{SIMT}{\glsdefaulttype}{SIMT}{single-instruction\, multiple-thread}{simt}
\myacronymentry{SPMD}{\glsdefaulttype}{SPMD}{single-program\, multiple-data}{spmd}

\myacronymentry{TAS}{\glsdefaulttype}{TAS}{test-and-set}{tas}
\myacronymentry{TCDM}{\glsdefaulttype}{TCDM}{tightly-coupled data memory}{tcdm}
\myacronymentry{TM}{\glsdefaulttype}{TM}{transactional memory}{tm}
\myacronymentry{Tx}{\glsdefaulttype}{Tx}{transmitter}{tx}

\myacronymentry{UART}{\glsdefaulttype}{UART}{universal asynchronous receiver transmitter}{uart}

\newcommand{\sy}{synchronization\xspace}

\renewcommand{\eqref}[1]{(\ref{#1})}    
\newcommand{\secref}[1]{\mbox{Sec.~\ref{#1}}}

\newcommand{\figref}[1]{\mbox{Fig.~\ref{#1}}}
\newcommand{\tblref}[1]{\mbox{Tbl.~\ref{#1}}}

\begin{document}
%
\newcommand{\ourTitle}{Energy-Efficient Hardware-Accelerated Synchronization for Shared-L1-Memory Multiprocessor Clusters}
\title{\ourTitle}
%
%
%
%

\author{Florian~Glaser,~\IEEEmembership{Student~Member,~IEEE,}
        Giuseppe~Tagliavini,~\IEEEmembership{Member,~IEEE,}
        Davide~Rossi,~\IEEEmembership{Member,~IEEE,}
        Germain~Haugou,
        Qiuting~Huang,~\IEEEmembership{Fellow,~IEEE,}
        and~Luca~Benini,~\IEEEmembership{Fellow,~IEEE}}

%
%

\markboth{IEEE Transactions on Parallel and Distributed Systems, vol.~XX, no.~xx, month~2020}{Glaser \MakeLowercase{\textit{et al.}}: \ourTitle}
\newcommand{\appRelCycleSCUmin}{1\%\xspace}
\newcommand{\appRelCycleSCUmax}{92\%\xspace}
\newcommand{\appRelCycleSCUavg}{23\%\xspace}
\newcommand{\appRelEnergySCUmin}{2\%\xspace}
\newcommand{\appRelEnergySCUmax}{98\%\xspace}
\newcommand{\appRelEnergySCUavg}{39\%\xspace}

\newcommand{\appRelCycleTASmin}{-9\%\xspace}
\newcommand{\appRelCycleTASmax}{3\%\xspace}
\newcommand{\appRelCycleTASavg}{-0\%\xspace}
\newcommand{\appRelEnergyTASmin}{-2\%\xspace}
\newcommand{\appRelEnergyTASmax}{65\%\xspace}
\newcommand{\appRelEnergyTASavg}{11\%\xspace}

\newcommand{\appIname}{DWT}
\newcommand{\appIdomain}{Signal processing}
\newcommand{\appInumBarr}{10}
\newcommand{\appIcodeSizeSCU}{xxx}
\newcommand{\appIcodeSizeTAS}{xxx}
\newcommand{\appIcodeSizeSW}{xxx}
\newcommand{\appIpowSCU}{21.7}
\newcommand{\appIpowTAS}{21.6}
\newcommand{\appIpowSW}{21.8}
\newcommand{\appISFRsizeSCU}{1.1k}
\newcommand{\appISFRsizeTAS}{1.1k}
\newcommand{\appISFRsizeSW}{1.1k}
\newcommand{\appIpowRedSCU}{0.77\%\xspace}
\newcommand{\appIpowRedTAS}{1.1\%\xspace}
\newcommand{\appIenergySCU}{0.7}
\newcommand{\appIenergyTAS}{0.8}
\newcommand{\appIenergySW}{0.8}
\newcommand{\appIexCycTotSCU}{11.3k}
\newcommand{\appIexCycTotTAS}{12.9k}
\newcommand{\appIexCycTotSW}{12.9k}
\newcommand{\appIexCycActSCU}{10.8k}
\newcommand{\appIexCycActTAS}{12.7k}
\newcommand{\appIexCycActSW}{12.9k}
\newcommand{\appIexCycActSprSCU}{155}
\newcommand{\appIexCycActSprTAS}{63}
\newcommand{\appIexCycActSprSW}{0}
\newcommand{\appIsyncCycTotSCU}{0.6k}
\newcommand{\appIsyncCycTotTAS}{1.5k}
\newcommand{\appIsyncCycTotSW}{1.6k}
\newcommand{\appIsyncCycActSCU}{84}
\newcommand{\appIsyncCycActTAS}{1.3k}
\newcommand{\appIsyncCycActSW}{1.6k}
\newcommand{\appIsyncCycTotRelSCU}{5.2\%}
\newcommand{\appIsyncCycTotRelTAS}{11.7\%}
\newcommand{\appIsyncCycTotRelSW}{12.6\%}
\newcommand{\appIsyncCycTotRedSCU}{2.8}
\newcommand{\appIsyncCycTotRedTAS}{1.1}
\newcommand{\appIsyncCycActRelSCU}{0.8\%}
\newcommand{\appIsyncCycActRelTAS}{10.0\%}
\newcommand{\appIsyncCycActRelSW}{12.7\%}
\newcommand{\appIsyncCycActRedSCU}{19}
\newcommand{\appIsyncCycActRedTAS}{1.3}
\newcommand{\appIIPCSCU}{5.01}
\newcommand{\appIIPCTAS}{4.65}
\newcommand{\appIIPCSW}{4.56}

\newcommand{\appIIname}{Dijkstra}
\newcommand{\appIIdomain}{Graph search}
\newcommand{\appIInumBarr}{238}
\newcommand{\appIIcodeSizeSCU}{xxx}
\newcommand{\appIIcodeSizeTAS}{xxx}
\newcommand{\appIIcodeSizeSW}{xxx}
\newcommand{\appIIpowSCU}{21.1}
\newcommand{\appIIpowTAS}{19.8}
\newcommand{\appIIpowSW}{21.4}
\newcommand{\appIISFRsizeSCU}{122}
\newcommand{\appIISFRsizeTAS}{156}
\newcommand{\appIISFRsizeSW}{130}
\newcommand{\appIIpowRedSCU}{1.4\%\xspace}
\newcommand{\appIIpowRedTAS}{7.4\%\xspace}
\newcommand{\appIIenergySCU}{2.0}
\newcommand{\appIIenergyTAS}{4.0}
\newcommand{\appIIenergySW}{4.0}
\newcommand{\appIIexCycTotSCU}{33.7k}
\newcommand{\appIIexCycTotTAS}{71.3k}
\newcommand{\appIIexCycTotSW}{64.9k}
\newcommand{\appIIexCycActSCU}{30.6k}
\newcommand{\appIIexCycActTAS}{69.1k}
\newcommand{\appIIexCycActSW}{64.9k}
\newcommand{\appIIexCycActSprSCU}{2.9k}
\newcommand{\appIIexCycActSprTAS}{0.7k}
\newcommand{\appIIexCycActSprSW}{0}
\newcommand{\appIIsyncCycTotSCU}{4.6k}
\newcommand{\appIIsyncCycTotTAS}{34.1k}
\newcommand{\appIIsyncCycTotSW}{34.0k}
\newcommand{\appIIsyncCycActSCU}{1.53k}
\newcommand{\appIIsyncCycActTAS}{32.0k}
\newcommand{\appIIsyncCycActSW}{34.0k}
\newcommand{\appIIsyncCycTotRelSCU}{13.7\%}
\newcommand{\appIIsyncCycTotRelTAS}{47.9\%}
\newcommand{\appIIsyncCycTotRelSW}{52.3\%}
\newcommand{\appIIsyncCycTotRedSCU}{7.3}
\newcommand{\appIIsyncCycTotRedTAS}{1}
\newcommand{\appIIsyncCycActRelSCU}{5.0\%}
\newcommand{\appIIsyncCycActRelTAS}{46.3\%}
\newcommand{\appIIsyncCycActRelSW}{52.3\%}
\newcommand{\appIIsyncCycActRedSCU}{22}
\newcommand{\appIIsyncCycActRedTAS}{1.1}
\newcommand{\appIIIPCSCU}{4.72}
\newcommand{\appIIIPCTAS}{4.48}
\newcommand{\appIIIPCSW}{4.09}

\newcommand{\appIIIname}{AES}
\newcommand{\appIIIdomain}{Cryptography}
\newcommand{\appIIInumBarr}{4}
\newcommand{\appIIIcodeSizeSCU}{xxx}
\newcommand{\appIIIcodeSizeTAS}{xxx}
\newcommand{\appIIIcodeSizeSW}{xxx}
\newcommand{\appIIIpowSCU}{23.8}
\newcommand{\appIIIpowTAS}{23.7}
\newcommand{\appIIIpowSW}{24.0}
\newcommand{\appIIISFRsizeSCU}{10.2k}
\newcommand{\appIIISFRsizeTAS}{10.2k}
\newcommand{\appIIISFRsizeSW}{10.2k}
\newcommand{\appIIIpowRedSCU}{1.1\%\xspace}
\newcommand{\appIIIpowRedTAS}{1.2\%\xspace}
\newcommand{\appIIIenergySCU}{2.8}
\newcommand{\appIIIenergyTAS}{2.8}
\newcommand{\appIIIenergySW}{2.9}
\newcommand{\appIIIexCycTotSCU}{41.2k}
\newcommand{\appIIIexCycTotTAS}{41.6k}
\newcommand{\appIIIexCycTotSW}{41.6k}
\newcommand{\appIIIexCycActSCU}{40.9k}
\newcommand{\appIIIexCycActTAS}{41.5k}
\newcommand{\appIIIexCycActSW}{41.6k}
\newcommand{\appIIIexCycActSprSCU}{188}
\newcommand{\appIIIexCycActSprTAS}{123}
\newcommand{\appIIIexCycActSprSW}{0}
\newcommand{\appIIIsyncCycTotSCU}{339}
\newcommand{\appIIIsyncCycTotTAS}{732}
\newcommand{\appIIIsyncCycTotSW}{719}
\newcommand{\appIIIsyncCycActSCU}{34}
\newcommand{\appIIIsyncCycActTAS}{547}
\newcommand{\appIIIsyncCycActSW}{719}
\newcommand{\appIIIsyncCycTotRelSCU}{0.8\%}
\newcommand{\appIIIsyncCycTotRelTAS}{1.8\%}
\newcommand{\appIIIsyncCycTotRelSW}{1.7\%}
\newcommand{\appIIIsyncCycTotRedSCU}{2.1}
\newcommand{\appIIIsyncCycTotRedTAS}{0.98}
\newcommand{\appIIIsyncCycActRelSCU}{0.1\%}
\newcommand{\appIIIsyncCycActRelTAS}{1.3\%}
\newcommand{\appIIIsyncCycActRelSW}{1.7\%}
\newcommand{\appIIIsyncCycActRedSCU}{21}
\newcommand{\appIIIsyncCycActRedTAS}{1.3}
\newcommand{\appIIIIPCSCU}{5.84}
\newcommand{\appIIIIPCTAS}{5.82}
\newcommand{\appIIIIPCSW}{5.80}

\newcommand{\appIVname}{Livermore6}
\newcommand{\appIVdomain}{Linear recurrence}
\newcommand{\appIVnumBarr}{127}
\newcommand{\appIVcodeSizeSCU}{xxx}
\newcommand{\appIVcodeSizeTAS}{xxx}
\newcommand{\appIVcodeSizeSW}{xxx}
\newcommand{\appIVpowSCU}{15.7}
\newcommand{\appIVpowTAS}{18.7}
\newcommand{\appIVpowSW}{22.0}
\newcommand{\appIVSFRsizeSCU}{104}
\newcommand{\appIVSFRsizeTAS}{104}
\newcommand{\appIVSFRsizeSW}{105}
\newcommand{\appIVpowRedSCU}{29\%\xspace}
\newcommand{\appIVpowRedTAS}{15\%\xspace}
\newcommand{\appIVenergySCU}{1.1}
\newcommand{\appIVenergyTAS}{1.7}
\newcommand{\appIVenergySW}{2.1}
\newcommand{\appIVexCycTotSCU}{24.5k}
\newcommand{\appIVexCycTotTAS}{32.3k}
\newcommand{\appIVexCycTotSW}{32.8k}
\newcommand{\appIVexCycActSCU}{14.0k}
\newcommand{\appIVexCycActTAS}{28.1k}
\newcommand{\appIVexCycActSW}{32.8k}
\newcommand{\appIVexCycActSprSCU}{6.8k}
\newcommand{\appIVexCycActSprTAS}{3.4k}
\newcommand{\appIVexCycActSprSW}{0}
\newcommand{\appIVsyncCycTotSCU}{11.3k}
\newcommand{\appIVsyncCycTotTAS}{19.1k}
\newcommand{\appIVsyncCycTotSW}{19.6k}
\newcommand{\appIVsyncCycActSCU}{760}
\newcommand{\appIVsyncCycActTAS}{14.9k}
\newcommand{\appIVsyncCycActSW}{19.6k}
\newcommand{\appIVsyncCycTotRelSCU}{46.1\%}
\newcommand{\appIVsyncCycTotRelTAS}{59.0\%}
\newcommand{\appIVsyncCycTotRelSW}{59.5\%}
\newcommand{\appIVsyncCycTotRedSCU}{1.7}
\newcommand{\appIVsyncCycTotRedTAS}{1}
\newcommand{\appIVsyncCycActRelSCU}{7.7\%}
\newcommand{\appIVsyncCycActRelTAS}{55.0\%}
\newcommand{\appIVsyncCycActRelSW}{59.5\%}
\newcommand{\appIVsyncCycActRedSCU}{26}
\newcommand{\appIVsyncCycActRedTAS}{1.3}
\newcommand{\appIVIPCSCU}{6.00}
\newcommand{\appIVIPCTAS}{5.25}
\newcommand{\appIVIPCSW}{4.74}

\newcommand{\appVname}{Livermore2}
\newcommand{\appVdomain}{Gradient descent}
\newcommand{\appVnumBarr}{12}
\newcommand{\appVcodeSizeSCU}{xxx}
\newcommand{\appVcodeSizeTAS}{xxx}
\newcommand{\appVcodeSizeSW}{xxx}
\newcommand{\appVpowSCU}{24.2}
\newcommand{\appVpowTAS}{22.9}
\newcommand{\appVpowSW}{23.4}
\newcommand{\appVSFRsizeSCU}{744}
\newcommand{\appVSFRsizeTAS}{789}
\newcommand{\appVSFRsizeSW}{788}
\newcommand{\appVpowRedSCU}{-3.7\%\xspace}
\newcommand{\appVpowRedTAS}{1.8\%\xspace}
\newcommand{\appVenergySCU}{0.6}
\newcommand{\appVenergyTAS}{0.7}
\newcommand{\appVenergySW}{0.8}
\newcommand{\appVexCycTotSCU}{9.2k}
\newcommand{\appVexCycTotTAS}{11.3k}
\newcommand{\appVexCycTotSW}{11.3k}
\newcommand{\appVexCycActSCU}{9.0k}
\newcommand{\appVexCycActTAS}{11.2k}
\newcommand{\appVexCycActSW}{11.3k}
\newcommand{\appVexCycActSprSCU}{46}
\newcommand{\appVexCycActSprTAS}{17}
\newcommand{\appVexCycActSprSW}{0}
\newcommand{\appVsyncCycTotSCU}{0.3k}
\newcommand{\appVsyncCycTotTAS}{1.8k}
\newcommand{\appVsyncCycTotSW}{1.8k}
\newcommand{\appVsyncCycActSCU}{71}
\newcommand{\appVsyncCycActTAS}{1.7k}
\newcommand{\appVsyncCycActSW}{1.8k}
\newcommand{\appVsyncCycTotRelSCU}{2.8\%}
\newcommand{\appVsyncCycTotRelTAS}{16.1\%}
\newcommand{\appVsyncCycTotRelSW}{16.1\%}
\newcommand{\appVsyncCycTotRedSCU}{7}
\newcommand{\appVsyncCycTotRedTAS}{1}
\newcommand{\appVsyncCycActRelSCU}{0.8\%}
\newcommand{\appVsyncCycActRelTAS}{15.4\%}
\newcommand{\appVsyncCycActRelSW}{16.1\%}
\newcommand{\appVsyncCycActRedSCU}{25}
\newcommand{\appVsyncCycActRedTAS}{1.1}
\newcommand{\appVIPCSCU}{6.67}
\newcommand{\appVIPCTAS}{5.94}
\newcommand{\appVIPCSW}{5.84}

\newcommand{\appVIname}{FFT}
\newcommand{\appVIdomain}{Frequency analysis}
\newcommand{\appVInumBarr}{4}
\newcommand{\appVIcodeSizeSCU}{xxx}
\newcommand{\appVIcodeSizeTAS}{xxx}
\newcommand{\appVIcodeSizeSW}{xxx}
\newcommand{\appVIpowSCU}{26.2}
\newcommand{\appVIpowTAS}{25.4}
\newcommand{\appVIpowSW}{26.0}
\newcommand{\appVISFRsizeSCU}{1.5k}
\newcommand{\appVISFRsizeTAS}{1.4k}
\newcommand{\appVISFRsizeSW}{1.4k}
\newcommand{\appVIpowRedSCU}{-0.83\%\xspace}
\newcommand{\appVIpowRedTAS}{2.3\%\xspace}
\newcommand{\appVIenergySCU}{0.5}
\newcommand{\appVIenergyTAS}{0.5}
\newcommand{\appVIenergySW}{0.5}
\newcommand{\appVIexCycTotSCU}{6.1k}
\newcommand{\appVIexCycTotTAS}{6.4k}
\newcommand{\appVIexCycTotSW}{6.4k}
\newcommand{\appVIexCycActSCU}{6.0k}
\newcommand{\appVIexCycActTAS}{6.3k}
\newcommand{\appVIexCycActSW}{6.4k}
\newcommand{\appVIexCycActSprSCU}{73}
\newcommand{\appVIexCycActSprTAS}{23}
\newcommand{\appVIexCycActSprSW}{0}
\newcommand{\appVIsyncCycTotSCU}{203}
\newcommand{\appVIsyncCycTotTAS}{606}
\newcommand{\appVIsyncCycTotSW}{670}
\newcommand{\appVIsyncCycActSCU}{39}
\newcommand{\appVIsyncCycActTAS}{540}
\newcommand{\appVIsyncCycActSW}{670}
\newcommand{\appVIsyncCycTotRelSCU}{3.3\%}
\newcommand{\appVIsyncCycTotRelTAS}{9.5\%}
\newcommand{\appVIsyncCycTotRelSW}{10.5\%}
\newcommand{\appVIsyncCycTotRedSCU}{3.3}
\newcommand{\appVIsyncCycTotRedTAS}{1.1}
\newcommand{\appVIsyncCycActRelSCU}{0.7\%}
\newcommand{\appVIsyncCycActRelTAS}{8.6\%}
\newcommand{\appVIsyncCycActRelSW}{10.5\%}
\newcommand{\appVIsyncCycActRedSCU}{17}
\newcommand{\appVIsyncCycActRedTAS}{1.2}
\newcommand{\appVIIPCSCU}{5.53}
\newcommand{\appVIIPCTAS}{5.40}
\newcommand{\appVIIPCSW}{5.34}

\newcommand{\appVIIname}{FANN-A}
\newcommand{\appVIIdomain}{Machine learning}
\newcommand{\appVIInumBarr}{160}
\newcommand{\appVIIcodeSizeSCU}{xxx}
\newcommand{\appVIIcodeSizeTAS}{xxx}
\newcommand{\appVIIcodeSizeSW}{xxx}
\newcommand{\appVIIpowSCU}{26.2}
\newcommand{\appVIIpowTAS}{26.2}
\newcommand{\appVIIpowSW}{26.5}
\newcommand{\appVIISFRsizeSCU}{519}
\newcommand{\appVIISFRsizeTAS}{483}
\newcommand{\appVIISFRsizeSW}{482}
\newcommand{\appVIIpowRedSCU}{0.96\%\xspace}
\newcommand{\appVIIpowRedTAS}{1.1\%\xspace}
\newcommand{\appVIIenergySCU}{6.9}
\newcommand{\appVIIenergyTAS}{7.7}
\newcommand{\appVIIenergySW}{7.9}
\newcommand{\appVIIexCycTotSCU}{92.4k}
\newcommand{\appVIIexCycTotTAS}{103.0k}
\newcommand{\appVIIexCycTotSW}{103.8k}
\newcommand{\appVIIexCycActSCU}{84.0k}
\newcommand{\appVIIexCycActTAS}{100.3k}
\newcommand{\appVIIexCycActSW}{103.8k}
\newcommand{\appVIIexCycActSprSCU}{2.2k}
\newcommand{\appVIIexCycActSprTAS}{0.9k}
\newcommand{\appVIIexCycActSprSW}{0}
\newcommand{\appVIIsyncCycTotSCU}{9.3k}
\newcommand{\appVIIsyncCycTotTAS}{25.7k}
\newcommand{\appVIIsyncCycTotSW}{26.7k}
\newcommand{\appVIIsyncCycActSCU}{982}
\newcommand{\appVIIsyncCycActTAS}{23.0k}
\newcommand{\appVIIsyncCycActSW}{26.7k}
\newcommand{\appVIIsyncCycTotRelSCU}{10.1\%}
\newcommand{\appVIIsyncCycTotRelTAS}{25.0\%}
\newcommand{\appVIIsyncCycTotRelSW}{25.8\%}
\newcommand{\appVIIsyncCycTotRedSCU}{2.9}
\newcommand{\appVIIsyncCycTotRedTAS}{1}
\newcommand{\appVIIsyncCycActRelSCU}{1.2\%}
\newcommand{\appVIIsyncCycActRelTAS}{23.0\%}
\newcommand{\appVIIsyncCycActRelSW}{25.8\%}
\newcommand{\appVIIsyncCycActRedSCU}{27}
\newcommand{\appVIIsyncCycActRedTAS}{1.2}
\newcommand{\appVIIIPCSCU}{6.72}
\newcommand{\appVIIIPCTAS}{6.48}
\newcommand{\appVIIIPCSW}{6.24}

\newcommand{\appVIIIname}{MFCC}
\newcommand{\appVIIIdomain}{Audio processing}
\newcommand{\appVIIInumBarr}{693}
\newcommand{\appVIIIcodeSizeSCU}{xxx}
\newcommand{\appVIIIcodeSizeTAS}{xxx}
\newcommand{\appVIIIcodeSizeSW}{xxx}
\newcommand{\appVIIIpowSCU}{23.8}
\newcommand{\appVIIIpowTAS}{22.8}
\newcommand{\appVIIIpowSW}{24.0}
\newcommand{\appVIIISFRsizeSCU}{718}
\newcommand{\appVIIISFRsizeTAS}{714}
\newcommand{\appVIIISFRsizeSW}{709}
\newcommand{\appVIIIpowRedSCU}{0.63\%\xspace}
\newcommand{\appVIIIpowRedTAS}{4.9\%\xspace}
\newcommand{\appVIIIenergySCU}{36.1}
\newcommand{\appVIIIenergyTAS}{41.5}
\newcommand{\appVIIIenergySW}{43.5}
\newcommand{\appVIIIexCycTotSCU}{0.53M}
\newcommand{\appVIIIexCycTotTAS}{0.64M}
\newcommand{\appVIIIexCycTotSW}{0.63M}
\newcommand{\appVIIIexCycActSCU}{0.50M}
\newcommand{\appVIIIexCycActTAS}{0.60M}
\newcommand{\appVIIIexCycActSW}{0.63M}
\newcommand{\appVIIIexCycActSprSCU}{14.8k}
\newcommand{\appVIIIexCycActSprTAS}{10.5k}
\newcommand{\appVIIIexCycActSprSW}{0}
\newcommand{\appVIIIsyncCycTotSCU}{33.1k}
\newcommand{\appVIIIsyncCycTotTAS}{142.3k}
\newcommand{\appVIIIsyncCycTotSW}{142.3k}
\newcommand{\appVIIIsyncCycActSCU}{4.64k}
\newcommand{\appVIIIsyncCycActTAS}{106.7k}
\newcommand{\appVIIIsyncCycActSW}{142.3k}
\newcommand{\appVIIIsyncCycTotRelSCU}{6.2\%}
\newcommand{\appVIIIsyncCycTotRelTAS}{22.3\%}
\newcommand{\appVIIIsyncCycTotRelSW}{22.4\%}
\newcommand{\appVIIIsyncCycTotRedSCU}{4.3}
\newcommand{\appVIIIsyncCycTotRedTAS}{1}
\newcommand{\appVIIIsyncCycActRelSCU}{0.9\%}
\newcommand{\appVIIIsyncCycActRelTAS}{17.8\%}
\newcommand{\appVIIIsyncCycActRelSW}{22.4\%}
\newcommand{\appVIIIsyncCycActRedSCU}{31}
\newcommand{\appVIIIsyncCycActRedTAS}{1.3}
\newcommand{\appVIIIIPCSCU}{6.84}
\newcommand{\appVIIIIPCTAS}{6.28}
\newcommand{\appVIIIIPCSW}{6.05}

\newcommand{\appIXname}{PCA}
\newcommand{\appIXdomain}{Data analysis}
\newcommand{\appIXnumBarr}{2305}
\newcommand{\appIXcodeSizeSCU}{xxx}
\newcommand{\appIXcodeSizeTAS}{xxx}
\newcommand{\appIXcodeSizeSW}{xxx}
\newcommand{\appIXpowSCU}{10.6}
\newcommand{\appIXpowTAS}{11.8}
\newcommand{\appIXpowSW}{19.0}
\newcommand{\appIXSFRsizeSCU}{375}
\newcommand{\appIXSFRsizeTAS}{388}
\newcommand{\appIXSFRsizeSW}{381}
\newcommand{\appIXpowRedSCU}{44\%\xspace}
\newcommand{\appIXpowRedTAS}{38\%\xspace}
\newcommand{\appIXenergySCU}{75.0}
\newcommand{\appIXenergyTAS}{89.6}
\newcommand{\appIXenergySW}{148.3}
\newcommand{\appIXexCycTotSCU}{2.48M}
\newcommand{\appIXexCycTotTAS}{2.66M}
\newcommand{\appIXexCycTotSW}{2.73M}
\newcommand{\appIXexCycActSCU}{0.88M}
\newcommand{\appIXexCycActTAS}{1.20M}
\newcommand{\appIXexCycActSW}{2.73M}
\newcommand{\appIXexCycActSprSCU}{0.6M}
\newcommand{\appIXexCycActSprTAS}{0.6M}
\newcommand{\appIXexCycActSprSW}{0}
\newcommand{\appIXsyncCycTotSCU}{1.62M}
\newcommand{\appIXsyncCycTotTAS}{1.76M}
\newcommand{\appIXsyncCycTotSW}{1.85M}
\newcommand{\appIXsyncCycActSCU}{20.55k}
\newcommand{\appIXsyncCycActTAS}{0.30M}
\newcommand{\appIXsyncCycActSW}{1.85M}
\newcommand{\appIXsyncCycTotRelSCU}{65.2\%}
\newcommand{\appIXsyncCycTotRelTAS}{66.3\%}
\newcommand{\appIXsyncCycTotRelSW}{67.8\%}
\newcommand{\appIXsyncCycTotRedSCU}{1.1}
\newcommand{\appIXsyncCycTotRedTAS}{1.1}
\newcommand{\appIXsyncCycActRelSCU}{2.9\%}
\newcommand{\appIXsyncCycActRelTAS}{29.6\%}
\newcommand{\appIXsyncCycActRelSW}{67.8\%}
\newcommand{\appIXsyncCycActRedSCU}{90}
\newcommand{\appIXsyncCycActRedTAS}{6.1}
\newcommand{\appIXIPCSCU}{4.47}
\newcommand{\appIXIPCTAS}{4.08}
\newcommand{\appIXIPCSW}{3.45}

\newcommand{\LivIICycleSampSCUmin}{2.6$\times$\xspace}
\newcommand{\LivIICycleSampSCUmax}{6.2$\times$\xspace}
\newcommand{\LivIICycleSampTASmin}{16\%\xspace}
\newcommand{\LivIICycleSampTASmax}{29\%\xspace}
\newcommand{\LivIICycleSampSWmin}{11\%\xspace}
\newcommand{\LivIICycleSampSWmax}{26\%\xspace}

\newcommand{\LivIICycleSartSCUmin}{26\%\xspace}
\newcommand{\LivIICycleSartSCUmax}{3.8$\times$\xspace}
\newcommand{\LivIICycleSartTASbet}{14\%\xspace}
\newcommand{\LivIICycleSartTASmin}{7\%\xspace}
\newcommand{\LivIICycleSartTASmax}{38\%\xspace}
\newcommand{\LivIICycleSartSWbet}{14\%\xspace}
\newcommand{\LivIICycleSartSWmin}{8\%\xspace}
\newcommand{\LivIICycleSartSWmax}{37\%\xspace}

\newcommand{\LivIICycleXiaoSCU}{8.9$\times$\xspace}
\newcommand{\LivIICycleXiaoTAS}{5$\times$\xspace}
\newcommand{\LivIICycleXiaoSW}{5$\times$\xspace}

\newcommand{\LivVICycleSampSCUmin}{13\%\xspace}
\newcommand{\LivVICycleSampSCUmax}{4.9$\times$\xspace}
\newcommand{\LivVICycleSampTASbet}{6\%\xspace}
\newcommand{\LivVICycleSampTASmin}{20\%\xspace}
\newcommand{\LivVICycleSampTASmax}{55\%\xspace}
\newcommand{\LivVICycleSampSWbet}{6\%\xspace}
\newcommand{\LivVICycleSampSWmin}{17\%\xspace}
\newcommand{\LivVICycleSampSWmax}{52\%\xspace}

\newcommand{\LivVICycleXiaoSCU}{3$\times$\xspace}
\newcommand{\LivVICycleXiaoTAS}{2.7$\times$\xspace}
\newcommand{\LivVICycleXiaoSW}{2.7$\times$\xspace}

\newcommand{\cycBarrSCUcoreI}{6}
\newcommand{\cycBarrSCUcoreII}{6}
\newcommand{\cycBarrSCUcoreIV}{6}
\newcommand{\cycBarrSCUcoreVIII}{6}

\newcommand{\cycBarrTAScoreI}{33}
\newcommand{\cycBarrTAScoreII}{52}
\newcommand{\cycBarrTAScoreIV}{91}
\newcommand{\cycBarrTAScoreVIII}{176}

\newcommand{\cycBarrSWcoreI}{27}
\newcommand{\cycBarrSWcoreII}{47}
\newcommand{\cycBarrSWcoreIV}{87}
\newcommand{\cycBarrSWcoreVIII}{176}

\newcommand{\cycleBarrRatioTAScoreI}{5.5}
\newcommand{\cycleBarrRatioTAScoreII}{8.7}
\newcommand{\cycleBarrRatioTAScoreIV}{15}
\newcommand{\cycleBarrRatioTAScoreVIII}{29}

\newcommand{\edgeSFRcycBarrSCUcoreI}{54}
\newcommand{\edgeSFRcycBarrSCUcoreII}{54}
\newcommand{\edgeSFRcycBarrSCUcoreIV}{54}
\newcommand{\edgeSFRcycBarrSCUcoreVIII}{54}

\newcommand{\edgeSFRcycBarrTAScoreI}{297}
\newcommand{\edgeSFRcycBarrTAScoreII}{468}
\newcommand{\edgeSFRcycBarrTAScoreIV}{819}
\newcommand{\edgeSFRcycBarrTAScoreVIII}{1584}

\newcommand{\edgeSFRcycBarrSWcoreI}{243}
\newcommand{\edgeSFRcycBarrSWcoreII}{423}
\newcommand{\edgeSFRcycBarrSWcoreIV}{783}
\newcommand{\edgeSFRcycBarrSWcoreVIII}{1584}

\newcommand{\edgeSFRcycleBarrRatioTAScoreI}{5.5}
\newcommand{\edgeSFRcycleBarrRatioTAScoreII}{8.7}
\newcommand{\edgeSFRcycleBarrRatioTAScoreIV}{15}
\newcommand{\edgeSFRcycleBarrRatioTAScoreVIII}{29}

\newcommand{\edgeSFRcycleBarrRatioSWcoreI}{4.5}
\newcommand{\edgeSFRcycleBarrRatioSWcoreII}{7.8}
\newcommand{\edgeSFRcycleBarrRatioSWcoreIV}{14}
\newcommand{\edgeSFRcycleBarrRatioSWcoreVIII}{29}

\newcommand{\energyBarrSCUcoreI}{0.1}
\newcommand{\energyBarrSCUcoreII}{0.1}
\newcommand{\energyBarrSCUcoreIV}{0.1}
\newcommand{\energyBarrSCUcoreVIII}{0.1}

\newcommand{\energyBarrTAScoreI}{0.5}
\newcommand{\energyBarrTAScoreII}{0.8}
\newcommand{\energyBarrTAScoreIV}{1.7}
\newcommand{\energyBarrTAScoreVIII}{4.3}

\newcommand{\energyBarrSWcoreI}{0.4}
\newcommand{\energyBarrSWcoreII}{0.8}
\newcommand{\energyBarrSWcoreIV}{1.8}
\newcommand{\energyBarrSWcoreVIII}{4.7}

\newcommand{\energyBarrRatioTAScoreI}{6.1}
\newcommand{\energyBarrRatioTAScoreII}{10}
\newcommand{\energyBarrRatioTAScoreIV}{18}
\newcommand{\energyBarrRatioTAScoreVIII}{38}

\newcommand{\energyBarrRatioSWcoreI}{4.9}
\newcommand{\energyBarrRatioSWcoreII}{9.4}
\newcommand{\energyBarrRatioSWcoreIV}{19}
\newcommand{\energyBarrRatioSWcoreVIII}{41}

\newcommand{\edgeSFRenergyBarrSCUcoreI}{51}
\newcommand{\edgeSFRenergyBarrSCUcoreII}{49}
\newcommand{\edgeSFRenergyBarrSCUcoreIV}{46}
\newcommand{\edgeSFRenergyBarrSCUcoreVIII}{42}

\newcommand{\edgeSFRenergyBarrTAScoreI}{313}
\newcommand{\edgeSFRenergyBarrTAScoreII}{503}
\newcommand{\edgeSFRenergyBarrTAScoreIV}{873}
\newcommand{\edgeSFRenergyBarrTAScoreVIII}{1622}

\newcommand{\edgeSFRenergyBarrSWcoreI}{252}
\newcommand{\edgeSFRenergyBarrSWcoreII}{464}
\newcommand{\edgeSFRenergyBarrSWcoreIV}{885}
\newcommand{\edgeSFRenergyBarrSWcoreVIII}{1771}

\newcommand{\edgeSFRenergyBarrRatioTAScoreI}{6.1}
\newcommand{\edgeSFRenergyBarrRatioTAScoreII}{10}
\newcommand{\edgeSFRenergyBarrRatioTAScoreIV}{18}
\newcommand{\edgeSFRenergyBarrRatioTAScoreVIII}{38}

\newcommand{\edgeSFRenergyBarrRatioSWcoreI}{4.9}
\newcommand{\edgeSFRenergyBarrRatioSWcoreII}{9.4}
\newcommand{\edgeSFRenergyBarrRatioSWcoreIV}{19}
\newcommand{\edgeSFRenergyBarrRatioSWcoreVIII}{41}

\newcommand{\cycMutexVSCUcoreI}{6}
\newcommand{\cycMutexVSCUcoreII}{12}
\newcommand{\cycMutexVSCUcoreIV}{23}
\newcommand{\cycMutexVSCUcoreVIII}{44}

\newcommand{\cycMutexVTAScoreI}{12}
\newcommand{\cycMutexVTAScoreII}{25}
\newcommand{\cycMutexVTAScoreIV}{39}
\newcommand{\cycMutexVTAScoreVIII}{69}

\newcommand{\cycMutexVSWcoreI}{5}
\newcommand{\cycMutexVSWcoreII}{12}
\newcommand{\cycMutexVSWcoreIV}{25}
\newcommand{\cycMutexVSWcoreVIII}{72}

\newcommand{\edgeSFRcycMutVSCUcoreI}{49}
\newcommand{\edgeSFRcycMutVSCUcoreII}{98}
\newcommand{\edgeSFRcycMutVSCUcoreIV}{187}
\newcommand{\edgeSFRcycMutVSCUcoreVIII}{356}

\newcommand{\edgeSFRcycMutVTAScoreI}{103}
\newcommand{\edgeSFRcycMutVTAScoreII}{215}
\newcommand{\edgeSFRcycMutVTAScoreIV}{331}
\newcommand{\edgeSFRcycMutVTAScoreVIII}{581}

\newcommand{\edgeSFRcycMutVSWcoreI}{40}
\newcommand{\edgeSFRcycMutVSWcoreII}{98}
\newcommand{\edgeSFRcycMutVSWcoreIV}{205}
\newcommand{\edgeSFRcycMutVSWcoreVIII}{608}

\newcommand{\edgeSFRcycleMutVratioTAScoreI}{2.1}
\newcommand{\edgeSFRcycleMutVratioTAScoreII}{2.2}
\newcommand{\edgeSFRcycleMutVratioTAScoreIV}{1.8}
\newcommand{\edgeSFRcycleMutVratioTAScoreVIII}{1.6}

\newcommand{\edgeSFRcycleMutVratioSWcoreI}{0.82}
\newcommand{\edgeSFRcycleMutVratioSWcoreII}{1}
\newcommand{\edgeSFRcycleMutVratioSWcoreIV}{1.1}
\newcommand{\edgeSFRcycleMutVratioSWcoreVIII}{1.7}

\newcommand{\energyMutexVSCUcoreI}{0.1}
\newcommand{\energyMutexVSCUcoreII}{0.2}
\newcommand{\energyMutexVSCUcoreIV}{0.3}
\newcommand{\energyMutexVSCUcoreVIII}{0.6}

\newcommand{\energyMutexVTAScoreI}{0.2}
\newcommand{\energyMutexVTAScoreII}{0.4}
\newcommand{\energyMutexVTAScoreIV}{0.7}
\newcommand{\energyMutexVTAScoreVIII}{1.6}

\newcommand{\energyMutexVSWcoreI}{0.1}
\newcommand{\energyMutexVSWcoreII}{0.2}
\newcommand{\energyMutexVSWcoreIV}{0.5}
\newcommand{\energyMutexVSWcoreVIII}{1.6}

\newcommand{\edgeSFRenergyMutVSCUcoreI}{50}
\newcommand{\edgeSFRenergyMutVSCUcoreII}{92}
\newcommand{\edgeSFRenergyMutVSCUcoreIV}{147}
\newcommand{\edgeSFRenergyMutVSCUcoreVIII}{211}

\newcommand{\edgeSFRenergyMutVTAScoreI}{112}
\newcommand{\edgeSFRenergyMutVTAScoreII}{231}
\newcommand{\edgeSFRenergyMutVTAScoreIV}{355}
\newcommand{\edgeSFRenergyMutVTAScoreVIII}{590}

\newcommand{\edgeSFRenergyMutVSWScoreI}{45}
\newcommand{\edgeSFRenergyMutVSWScoreII}{114}
\newcommand{\edgeSFRenergyMutVSWScoreIV}{239}
\newcommand{\edgeSFRenergyMutVSWScoreVIII}{574}

\newcommand{\edgeSFRenergyMutVratioTAScoreI}{2.2}
\newcommand{\edgeSFRenergyMutVratioTAScoreII}{2.5}
\newcommand{\edgeSFRenergyMutVratioTAScoreIV}{2.4}
\newcommand{\edgeSFRenergyMutVratioTAScoreVIII}{2.8}

\newcommand{\edgeSFRenergyMutVratioSWcoreI}{0.91}
\newcommand{\edgeSFRenergyMutVratioSWcoreII}{1.2}
\newcommand{\edgeSFRenergyMutVratioSWcoreIV}{1.6}
\newcommand{\edgeSFRenergyMutVratioSWcoreVIII}{2.7}

\newcommand{\cycMutexXSCUcoreI}{7}
\newcommand{\cycMutexXSCUcoreII}{13}
\newcommand{\cycMutexXSCUcoreIV}{24}
\newcommand{\cycMutexXSCUcoreVIII}{50}

\newcommand{\cycMutexXTAScoreI}{14}
\newcommand{\cycMutexXTAScoreII}{26}
\newcommand{\cycMutexXTAScoreIV}{50}
\newcommand{\cycMutexXTAScoreVIII}{89}

\newcommand{\cycMutexXSWcoreI}{6}
\newcommand{\cycMutexXSWcoreII}{13}
\newcommand{\cycMutexXSWcoreIV}{26}
\newcommand{\cycMutexXSWcoreVIII}{55}

\newcommand{\edgeSFRcycMutXSCUcoreI}{53}
\newcommand{\edgeSFRcycMutXSCUcoreII}{97}
\newcommand{\edgeSFRcycMutXSCUcoreIV}{176}
\newcommand{\edgeSFRcycMutXSCUcoreVIII}{370}

\newcommand{\edgeSFRcycMutXTAScoreI}{116}
\newcommand{\edgeSFRcycMutXTAScoreII}{214}
\newcommand{\edgeSFRcycMutXTAScoreIV}{410}
\newcommand{\edgeSFRcycMutXTAScoreVIII}{721}

\newcommand{\edgeSFRcycMutXSWcoreI}{44}
\newcommand{\edgeSFRcycMutXSWcoreII}{97}
\newcommand{\edgeSFRcycMutXSWcoreIV}{194}
\newcommand{\edgeSFRcycMutXSWcoreVIII}{415}

\newcommand{\edgeSFRcycleMutXratioTAScoreI}{2.2}
\newcommand{\edgeSFRcycleMutXratioTAScoreII}{2.2}
\newcommand{\edgeSFRcycleMutXratioTAScoreIV}{2.3}
\newcommand{\edgeSFRcycleMutXratioTAScoreVIII}{1.9}

\newcommand{\edgeSFRcycleMutXratioSWcoreI}{0.83}
\newcommand{\edgeSFRcycleMutXratioSWcoreII}{1}
\newcommand{\edgeSFRcycleMutXratioSWcoreIV}{1.1}
\newcommand{\edgeSFRcycleMutXratioSWcoreVIII}{1.1}

\newcommand{\energyMutexXSCUcoreI}{0.1}
\newcommand{\energyMutexXSCUcoreII}{0.2}
\newcommand{\energyMutexXSCUcoreIV}{0.3}
\newcommand{\energyMutexXSCUcoreVIII}{0.7}

\newcommand{\energyMutexXTAScoreI}{0.2}
\newcommand{\energyMutexXTAScoreII}{0.4}
\newcommand{\energyMutexXTAScoreIV}{0.9}
\newcommand{\energyMutexXTAScoreVIII}{2.1}

\newcommand{\energyMutexXSWcoreI}{0.1}
\newcommand{\energyMutexXSWcoreII}{0.2}
\newcommand{\energyMutexXSWcoreIV}{0.6}
\newcommand{\energyMutexXSWcoreVIII}{1.5}

\newcommand{\edgeSFRenergyMutXSCUcoreI}{55}
\newcommand{\edgeSFRenergyMutXSCUcoreII}{91}
\newcommand{\edgeSFRenergyMutXSCUcoreIV}{139}
\newcommand{\edgeSFRenergyMutXSCUcoreVIII}{221}

\newcommand{\edgeSFRenergyMutXTAScoreI}{125}
\newcommand{\edgeSFRenergyMutXTAScoreII}{232}
\newcommand{\edgeSFRenergyMutXTAScoreIV}{429}
\newcommand{\edgeSFRenergyMutXTAScoreVIII}{738}

\newcommand{\edgeSFRenergyMutXSWScoreI}{50}
\newcommand{\edgeSFRenergyMutXSWScoreII}{123}
\newcommand{\edgeSFRenergyMutXSWScoreIV}{269}
\newcommand{\edgeSFRenergyMutXSWScoreVIII}{510}

\newcommand{\edgeSFRenergyMutXratioTAScoreI}{2.3}
\newcommand{\edgeSFRenergyMutXratioTAScoreII}{2.5}
\newcommand{\edgeSFRenergyMutXratioTAScoreIV}{3.1}
\newcommand{\edgeSFRenergyMutXratioTAScoreVIII}{3.3}

\newcommand{\edgeSFRenergyMutXratioSWcoreI}{0.92}
\newcommand{\edgeSFRenergyMutXratioSWcoreII}{1.3}
\newcommand{\edgeSFRenergyMutXratioSWcoreIV}{1.9}
\newcommand{\edgeSFRenergyMutXratioSWcoreVIII}{2.3}

\IEEEtitleabstractindextext{%
\begin{abstract}

The steeply growing performance demands for highly power- and energy-constrained processing systems such as end-nodes of the internet-of-things (IoT) have led to parallel near-threshold computing (NTC), joining the energy-efficiency benefits of low-voltage operation with the performance typical of parallel systems. Shared-L1-memory multiprocessor clusters are a promising architecture, delivering performance in the order of GOPS and over 100\,GOPS/W of energy-efficiency. However, this level of computational efficiency can only be reached by maximizing the effective utilization of the processing elements (PEs) available in the clusters. Along with this effort, the optimization of PE-to-PE \sy and communication is a critical factor for performance. 
In this work, we describe a light-weight hardware-accelerated \sy and communication unit (SCU) for tightly-coupled clusters of processors. We detail the architecture, which enables fine-grain per-PE power management, and its integration into an eight-core cluster of RISC-V processors.
To validate the effectiveness of the proposed solution, we implemented the eight-core cluster in advanced 22\,nm FDX technology and evaluated performance and energy-efficiency with tunable microbenchmarks and a set of real-life applications and kernels. The proposed solution allows \sy-free regions as small as \edgeSFRenergyBarrSCUcoreVIII\xspace cycles, over \edgeSFRenergyBarrRatioSWcoreVIII$\times$ smaller than the baseline implementation based on fast test-and-set access to L1 memory when constraining the microbenchmarks to 10\% \sy overhead. When evaluated on the real-life DSP-applications, the proposed SCU improves performance by up to \appRelCycleSCUmax and \appRelCycleSCUavg on average and energy efficiency by up to \appRelEnergySCUmax and \appRelEnergySCUavg on average.

\end{abstract}

}

\maketitle

\IEEEdisplaynontitleabstractindextext

%
\IEEEpeerreviewmaketitle

\IEEEraisesectionheading{\section{Introduction}\label{sec:intro}}

%
%
%
%

\IEEEPARstart{A}{fter} being established as the architectural standard for general-purpose and high-performance computing over a decade ago \cite{Genbrugge09,Geer05}, the paradigm of \glspl{CMP} has as well been adopted in the embedded computing domain \cite{Bertozzi05,Levy09}. While the main driving force for the former domain is prohibitive heat dissipation as a result of ever-rising clock frequencies of up to several GHz, the turn to parallel processing in the latter domain is propelled by the trend toward high computational performance without losing the energy efficiency of the so far used simple and low-performance microcontrollers. A key contributor to the increase in performance demand in embedded devices is the rise of \gls{IoT} end-nodes that need to flexibly handle multiple sensor data streams \cite{Khan12,Gubbi13} (e.g., from low-resolution cameras or microphone arrays) and perform complex computations on them to reduce the bandwidth over energy-intensive wireless data links.

As the straightforward replacement of microcontroller cores with more powerful core variants featuring multiple-issue, multiple-data pipeline stages, and higher operating frequencies, naturally jeopardizes energy efficiency \cite{AziziISCA2010}, researchers are turning to \textit{parallel} \gls{NTC}. Reducing the supply voltage of the underlying CMOS circuits to their \gls{OEP} \cite{Salamin19}, usually located slightly above their threshold voltage, enables improvements of energy efficiency by up to one order of magnitude \cite{Dreslinski10}. The gain in energy efficiency, however, comes with a significant loss in performance of around 10$\times$ \cite{Dreslinski10} due to the reduced maximum operating frequency directly linked to voltage scaling. To overcome this performance limitation, we resort to parallel \gls{NTC} \cite{PULPv2}, an approach coupling the high energy efficiency of near-threshold operation with the performance typical of tightly coupled clusters of \glspl{PE} that can, when utilized in parallel, recover from the supply-scaling induced performance losses, delivering up to GOPS in advanced technology nodes. However, parallel \gls{NTC} can only achieve the fundamental goal of increasing energy efficiency with parallel workloads and, when the utilization of computational resources is well balanced, or, more in general, when the underlying hardware can effectively exploit the parallelism present in applications.

If this is not the case, performance loss must be recovered by increasing the operating frequency (and the supply voltage) to achieve the given target, thereby reducing the energy efficiency of the system. Moreover, in sequential portions of applications, parallel hardware resources such as \glspl{PE} and part of the interconnect towards the shared memory consume power without contributing to performance. Hence, in these regions, all idle components must be aggressively power-managed at a fine-grain level. 

The described requirements highlight the need for communication, \sy, and power management support in parallel clusters.
Communication mechanisms allow \glspl{PE} to communicate with each other to exchange intermediate results and orchestrate parallel execution. In this work, we focus on \textit{shared-memory} multiprocessors that typically rely on data-parallel computational models. For this class of systems, data exchange is trivial, restricting the communication aspects to pointer exchange and data validity signaling. However, wait-and-notify primitives are required by every application that has any form of data dependency between threads (i.e., that cannot be \textit{vectorized}). Consequently, the support for \sy mechanisms remains mandatory also for this type of parallel processing systems.

The most straightforward way to enable functionally correct implementations of every kind of multi-\gls{PE} \sy is to provide atomic accesses to the shared memory (or a part thereof) and use \textit{spinlocks} on protected shared variables. While this approach is universal, flexible, and requires small hardware overhead, it constitutes a form of \textit{busy-waiting} as every contestant repeatedly accesses the shared variables until all get exclusive access (at least once, depending on the \sy primitive that is implemented). This concept is not only prohibitive from an energy efficiency viewpoint due to the wasted energy for every failing lock-acquire attempt but is also considerably disadvantageous from a performance point-of-view as the concurrent attempts can put high loads on processor-to-memory interconnect systems and cause contentions on the shared memory system. Additionally, as every contestant has to acquire the lock sequentially, the cost of \sy primitives in terms of cycles is even in the best case lower bounded by the product of memory access latency and the number of involved \glspl{PE}, therefore growing with the number of contestants.

Basic interrupt and \gls{PM} support allow avoiding busy-waiting with the help of software \sy primitives that only require all contestants to become active after updating the shared atomic variables resp. to change spin lock ownership. While the context changes required for interrupt handling naturally incur software overheads, the issues associated with concurrent lock-acquire attempts, additionally, the cost and scaling of \sy primitives remain unaffected. As a result, the handling of a single \sy point can take over a hundred cycles even with less than ten involved \glspl{PE} as our experiments in \secref{sec:synth_bench} show. With the constraint to gain from parallelization, either in terms of energy or performance or both, this cost figure causes a lower bound for the average length of periods that \glspl{PE} can work independently from each other, often referred to as the \textit{\sy-free region}. In this work, we use the term \textit{parallel section} interchangeably, as it is often found in the context of parallel programming models.

The fundamental principle in parallel \gls{NTC} of utilizing all available computational resources as equally as possible conflicts heavily with a minimum required \sy-free region length: If parallelizing an application only pays off for parallel sections of thousands of cycles or more due to the associated \sy overhead, energy-efficient execution of all applications that exhibit finer-grain inter-thread dependencies is thwarted even though the overall system architecture would be perfectly suitable.

To overcome the explained challenges and limitations, we propose a light-weight hardware-supported solution that aims at drastically reducing the \sy overhead in terms of cycles and -- more importantly -- energy, thereby making fine-grain parallelization for the targeted shared-memory \gls{NTC} processing clusters affordable. All signaling is done by \textit{restfully waiting} on \textit{events}, i.e., halting and resuming execution at \sy points without the need to change the software context. We detail the hardware architecture of the \gls{SCU}, the foundation of the proposed solution that centrally manages event generation, and is instantiated as a single-cycle accessible shared peripheral. It provides both general-purpose signaling and an easily extensible set of commonly used \sy primitives. In this work, we focus on the \textit{barrier} and \textit{mutex} primitives, as they are required for the parallel and critical section constructs that are fundamental in parallel programming frameworks such as OpenMP \cite{OpenMP}.

For cases where a completely balanced utilization of all \glspl{PE} is not possible, we propose fine-grain \gls{PM} in the form of clock gating, fused into the \gls{SCU}, that allows saving energy during idle periods as short as tens of cycles. To demonstrate the capabilities of the solution, we integrate the \gls{SCU} into an eight-core heterogeneous cluster as an example for the targeted systems and how it can be efficiently used from a software point-of-view by extending the ISA of the \gls{DSP}-enabled RISC-V cores with only a single instruction. We illustrate both the opportunities and relevance of lowering \sy overhead for parallel \gls{NTC} through synthetic benchmarks and a set of \gls{DSP} kernels that are typical for the targeted system type. By analyzing the performance and energy of the whole system in both cases, we demonstrate not only the theoretically possible but also the practical gains of the proposed solution. As a competitive baseline, we use software implementations of the primitives based on atomic L1 memory access in the form of \gls{TAS}; for a fair comparison, the baseline implementations also include variants that employ event-based restful waiting. To achieve reliable results in the context of fine-grain parallelization, we carry out all experiments on a gate-level, fabrication-ready implementation of the multicore cluster in a 22\,nm process, allowing us to obtain cycle-exact performance numbers as well as to measure energy with an accuracy close to that of physical system realizations. The results obtained in our evaluation show that the \gls{SCU} allows \sy-free regions as small as \edgeSFRenergyBarrSCUcoreVIII\xspace cycles, which is more than \edgeSFRenergyBarrRatioSWcoreVIII$\times$ smaller than the implementation based on fast \gls{TAS} when constraining the synthetic benchmarks to 10\% synchronization overhead. Moreover, when evaluated on real-life kernels, the proposed SCU improves performance by up to \appRelCycleSCUmax and energy efficiency by up to \appRelEnergySCUmax.

The remainder of this paper is organized as follows. \secref{sec:relwork} provides a comprehensive overview of the prior art related to \sy in embedded multiprocessors. \secref{sec:background} discusses the relevance of fine-grain \sy in the context of the targeted system type. The architecture of both the proposed \gls{SCU} and the hosting multiprocessor is explained in \secref{sec:archi}, followed by the thereby enabled concept of aggressively reducing the overhead for \sy primitives in \secref{sec:single_sync}. The baseline method, as well as the experimental setup and methodology which we used, can be found in \secref{sec:exp}, followed by the experimental results for both the microbenchmarks and the range of \gls{DSP}-applications. \secref{sec:conclusion} summarizes and concludes the work.

\section{Related Work}\label{sec:relwork}

The shortcomings associated with straight-forward \sy support based on atomic memory access have been broadly recognized by the research community; a variety of works, therefore, proposes, similar to our approach, hardware-accelerated solutions to improve performance \cite{DeMatteis13,Sartori10,Sampson06,Xiao12,Saglam01,Beckmann90}, energy efficiency \cite{Ferri08,Kim14}, or both \cite{Monchiero06,Yu10}. Reviews of the performance and characteristics of software-based solutions for shared-memory multiprocessors can be found in \cite{Anderson90,Kaegi97,Mellor91,Ferri09,Golubeva07} and date back as early as the 1990s. Multiple works \cite{Tsai16,Ferri09,Loghi06,Liu05} propose to dynamically adjust the speed of individual \glspl{PE} at runtime to equalize their execution speed instead of power-managing them. This approach, however, incurs significant control and circuit complexity due to the required asynchronous clocks and severely increases the latency between \glspl{PE} and memory as crossing clock domain boundaries already takes several cycles. Consequently, it only pays off if the costs for entering and leaving low-power modes are in the order of thousands of cycles, as the authors of \cite{Ferri09} assume. Our approach instead aims at using a single synchronous clock, implementing a simple variant of \gls{PM} that is suitable for very short idle periods, equalizing workload with fine-grain parallelization and ultimately gain from system-wide frequency and supply scaling.

The prior art in hardware-accelerated \sy for embedded, synchronously clocked systems covers a wide range of target architectures and implementation concepts, is, however, to a large extent, not suitable for the microcontroller-class cacheless shared memory type of system that we target. What follows is a review of the corresponding references, structured by key aspects that illustrate the causes for the stated mismatch.

\textbf{Synchronization-free region size:}
As explained previously in \secref{sec:intro}, the ability to handle typical \sy tasks in the order of tens of cycles and below is of major importance for energy-efficient parallelization in the context of the targeted processing clusters. Except for \cite{Beckmann90}, no explicit statement is made about the targeted \gls{SFR} size. The authors of \cite{Beckmann90} report the speedup for \glspl{SFR} as small as ten cycles; however, leave many implementation details open. For all other references, the \gls{SFR} size for which the respective results are reported must be implicitly determined by analyzing the employed benchmarks. Multiple references \cite{Monchiero06,Kim14,Yu10} employ rather dated parallel benchmarking suites such as STAMP \cite{STAMP} or SPLASH-2 \cite{SPLASH2} which feature \glspl{SFR} of 10.000s of cycles and are, therefore -- in addition to prohibitive data set sizes -- not suitable for the microcontroller-type clusters we target. The order of magnitude for the \gls{SFR} size of the mentioned benchmarks was determined through experiments on a general-purpose desktop computer; the misfit of, e.g., the SPLASH-2 suite, is also mentioned in \cite{Sampson06}. As a suitable alternative, the authors of \cite{Sampson06} propose the usage of a subset of the Livermore Loops \cite{Feo88}, a collection of sequential \gls{DSP}-kernels that, however, can be parallelized with reasonable effort as the original code is annotated with data hazards and the like. The parallel versions of loops 2,3, and 6 are provided in \cite{Sampson06} and used in \cite{Sampson06,Xiao12,Sartori10}; we include loops 2 and 6 in our set of applications; loop 3 is omitted as it is a fully vectorizable matrix multiplications and \sy consequently not required within the kernel. Our analysis in \secref{sec:exp} verifies the fitting \gls{SFR} sizes of loops 2 and 6; they are as small as \appIVSFRsizeSCU\,cycles. Whenever provided, we compare the execution time for both loops achieved with our solution to the references in question and observe better performance even when considering systems with higher \gls{PE} counts as ours.

In addition to the conclusions drawn from the used benchmarks, a closer analysis of the employed bus systems can also help to estimate the smallest supported \gls{SFR}: For example, the \gls{SB} proposed in \cite{Monchiero06}, managing all \sy constructs locally at the shared memory, is only reachable for the \glspl{PE} through a \gls{NoC}. A latency of at least 30\,cycles for \gls{NoC} transactions is stated; for requesting and getting notified about the availability of a lock, at least two transactions are required, pushing the affordable \gls{SFR} size far above the approximately one hundred cycles and below targeted by us.

The authors of \cite{Ferri08,Yu10,Saglam01} either explicitly state task-level parallelism or use benchmarks that feature software pipelining (of independent algorithm parts) and therefore do not target \sy at loop-level of whatever granularity.

Due to the similarities in the targeted systems, it can be assumed that \cite{Thabet13} aims at similarly sized \sy-free regions as ours, although not explicitly stated. Unfortunately, results are only reported as number and types of accesses to the \sy hardware for various \sy primitives. The \textit{barrier} and \textit{mutex} primitives that we discuss in detail are estimated to take two bus transactions or six cycles (measured at the memory bus); we reduce the number of bus transactions to one and the latency to four cycles; additionally, we include the feedback of primitive-specific information.

\textbf{Synchronization hardware complexity:}
Keeping overall circuit area and complexity as small as possible is of crucial importance for systems that ought to be employed for parallel \gls{NTC}, as explained in \secref{sec:intro}. Naturally, this also holds true for any \sy-managing unit that must both affect overall area only insignificantly and also not introduce complex (possibly latency-critical) circuitry due to the associated dynamic power, thereby diminishing the savings obtained through the acceleration of \sy tasks.

The cache-alike \gls{SB} presented in \cite{Monchiero06} does not meet these requirements; considerable circuit area is required for each entry to store all required information as well as for the logic that performs single-cycle hit-detection. The need to check every memory access for a matching address causes the activity of the \gls{SB} to be much higher than the frequency of \sy points in a given application would require.

The concepts presented in \cite{Ferri08} and \cite{Yu10} are based on snoop devices at the memory, or system bus; \cite{Yu10} is of particular interest as it distributes the \sy-management over one controller per \gls{PE} that each hosts a locking queue for the respective variables of interest. This concept may sound very appealing from a hardware complexity point of view; however, an important aspect not covered in \cite{Yu10} changes the picture significantly: In the usual (and desirable) case where the system bus of a \gls{CMP} can handle multiple transactions at once, each local \sy controller must have global visibility of the bus and be able to \textit{parallelly} check the maximum number of concurrent transactions against \textit{any} of its monitored locking variables. It is obvious that even a single such device cannot be built in a slim way, replicating it for every \gls{PE} makes the situation worse.

Although no analysis is provided, the complexity of the barrier filter modules proposed in \cite{Sampson06} can be estimated to be slightly higher than for our proposed hardware barrier modules (based on the available knowledge about the amount of information that each barrier filter needs to store). We reduce all address-related housekeeping overhead and restrictions by assigning each barrier module a fixed address in the global peripheral address space of the system and by providing \gls{PE}-parallel access.

This work is most comparable to the hardware synchronizer (HWS) proposed in \cite{Thabet13} that is connected as a shared peripheral, enabling \sy primitives through appropriate programming of an array of atomic counters and compare registers. We reduce the hardware complexity and remove the burden of configuring and mapping the atomic counters to \sy primitives by providing native and low-cost hardware support for such while maintaining general-purpose signaling.

The concepts presented in \cite{Saglam01,Beckmann90} principally match ours as shared, dedicated registers are used to represent the state of \sy primitives at low hardware cost. Sadly, many implementation and integration aspects are left undiscussed, and either only locks \cite{Saglam01} or only barriers \cite{Beckmann90} are natively supported. While a hardware lock can be employed to realize a barrier, the cycle cost compared to a hardware barrier is clearly prohibitive for the small \gls{SFR} sizes that we target. While our solution also features general-purpose atomic \gls{PE}-to-\gls{PE} signaling, the most critical feature is native hardware support for the most commonly used \sy primitives (in the targeted systems and programming models) as well as concurrent access to them at very small hardware overhead.

\textbf{Target system type:}
The vast majority of prior art assumes processing systems with data caches; those either affect the respectively proposed memory-mapped \sy concept or are even actively modified and used for \sy \cite{Sampson06}. The exceptions \cite{Yu10,Thabet13,Saglam01,Beckmann90} propose \sy solutions that sit closer to the \glspl{PE} than any data cache or are too vague in the description of the targeted systems to decide about the existence of caches.

The implications of the targeted system on the complexity of the \sy hardware are well illustrated in, e.g., \cite{Monchiero06,Sampson06,Xiao12} that all target high-end architectures with much higher operating frequencies than we do, at the cost of sacrificing energy efficiency. To ensure scalability beyond a few \glspl{PE} as well as the high operating frequencies, all shared memory and peripherals are only reachable through high-latency buses or networks. As any \sy-managing unit must conceptually have global observability of \sy requests, it must be either put in front of the shared memory and keep track of (i.e., \textit{store}) the sequential accesses to such \cite{Monchiero06,Sampson06}, or better suitable, dedicated message-passing networks must be added \cite{Xiao12}; both alternatives infer undesirable complex hardware. In the latter case, the primary focus is on latency and bandwidth of \gls{PE}-to-\gls{PE} transfers while the \sy functionality is conveniently added but would not require the underlying high-bandwidth message-passing hardware. The low-latency and concurrent access to shared memory in our targeted clusters allow us to limit the \sy hardware to a \textit{signaling} role and use the shared memory for any data exchange that is larger than a single word without performance penalties.

On the other side of the spectrum, popular shared-memory parallel architectures are \glspl{GPU}, employing the \gls{SIMT} execution model, a combination of \gls{SIMD} and multithreading. \Glspl{GP-GPU} are hierarchical architectures composed of multiple clusters (called multiprocessors) that, in turn, are made up of multiple \glspl{PE}. Each \gls{PE} executes a group of threads (called a warp) in lockstep, according to the order of instructions issued by a dispatcher, which is shared among all of them \cite{hennessy_patterson_5th}. While in traditional \glspl{GP-GPU}, only global synchronization primitives and hardware support were available, only allowing to synchronize all threads running on a multiprocessor, recent architectures such as Nvidia Volta allow to synchronize the threads within a warp, improving the synchronization efficiency for kernels with smaller granularity \cite{choquette2018volta}. However, the \gls{SIMT} nature of \glspl{GP-GPU} forces them to sequentialize divergent threads and threads with critical sections, jeopardizing performance and energy efficiency, which makes them very inflexible and definitively not suitable for the application domain targeted in this work.

A hybrid \gls{SIMD}/MIMD (multiple-instruction, multiple-data) approach has been proposed in \cite{epfl}. The architecture combines a tightly coupled cluster of processors featuring a shared instruction memory with broadcast capability and a counter-based hardware synchronization mechanism, dynamically managing the lockstep execution of cores during data-dependent program flows. Although this approach achieves 60\% of energy reduction, its applicability is restricted to data-parallel code sequences, and it is intrusive from the software viewpoint, as it requires explicitly managing counters and lockstep execution. On the other hand, the approach proposed in this work is fully flexible and provides easily usable primitives supported by parallel programming models such as OpenMP \cite{OpenMP}.

\textbf{\gls{PM} and signaling mechanism:} With the exception of \cite{Monchiero06,Sampson06,Sartori10}, where no explicit statements are made, all references implement or at least suggest idle waiting for \glspl{PE} that are blocked at \sy points. The majority of the works with a focus on idle waiting proposes interrupt-based mechanisms \cite{Ferri08,Thabet13,Saglam01,Beckmann90} without further consideration for the associated context switching overhead. Event-based signaling is supported in \cite{Thabet13}; however, no details are given on how \gls{PM} and idle-waiting are realized. The authors of \cite{Braojos17,Kim14,Xiao12} follow our approach of power-managing idle cores through clock gating; power and area of the section-monitoring pool units in \cite{Kim14} are, however, not suitable for our microcontroller-class target systems. The architecture of the system proposed in \cite{Braojos17} is similar to ours; multiple \glspl{PE}, operating in lockstep, are connected to a multi-banked memory through a single-cycle crossbar. However, the targeted lockstep operation of the \glspl{PE} greatly limits the generality of how the system can be used. Furthermore, no details about the architecture and integration of the hardware \sy unit (SU), that is central to the design, are provided. 

In contrast, \cite{Yu10,Sampson06} propose to \textit{stall} \glspl{PE} that cannot continue by means of absent replies to requests on their data- or instruction ports. As this approach allows handling the aforementioned check and decision for continuation in one operation as well as {PE}-externally and centrally, we follow it in favor of reducing the time devoted to \sy.
The solution proposed in \cite{Thabet13} is the only one that combines low cycle-overhead \sy with \gls{PM}; unfortunately, an analysis of the achievable gains in terms of system performance and energy efficiency is not provided. We combine the concepts of stalling cores that are blocked at \sy points and event-based signaling, and tightly couple those with per-core fine-grain clock gating to allow the handling of \sy primitives in less than ten active \gls{PE} cycles. Large parts of the required hardware are reused to also provide general-purpose interrupt support for, e.g., the handling of data exceptions.

\textbf{Implementation stage:} The majority of the previous works employs behavioral models of the individual system components (\glspl{PE}, interconnect, memories, \sy hardware) written in higher-level languages and instruction- or transaction-level simulators such as, e.g., GRAPES in \cite{Monchiero06}, MPARM in \cite{Kim14,Ferri08}, or M5 in \cite{Yu10,Sartori10}. While this approach enables the simulation of complex and large-scale architectures in reasonable times, it has the drawback of reduced accuracy for the figures of interest, performance (or execution time) and power, when compared to cycle-exact simulations based on synthesizable modules captured in a \gls{HDL} or gate-level implementations. The loss in accuracy may be acceptable for evaluating the performance of \sy solutions for task-level parallelism (where the main goal is to reduce or eliminate polling over high-performance interconnect systems) or the support of \gls{TM} with applications that spend 50\% and more of their execution time within critical sections \cite{Kim14}. For our goal of enabling fine-grain parallelism with few tens or hundreds of cycles between \sy points, however, cycle-level accuracy is required to reliably evaluate the effects of different solutions with an increasing degree of hardware support (from atomic memory access to full \sy primitives). For example, slight differences in the arrival instants of \glspl{PE} at \sy points due to, e.g., cache misses or small workload imbalances can have a massive impact on the subsequently caused contention during lock acquire trials, as we demonstrate in \secref{sec:exp}.

To the best of the authors' knowledge, \cite{Thabet13} is the only prior work that reports gate-level (and even silicon) implementations of the whole processing system, including the \sy hardware. Unfortunately, only latencies for various \sy primitives in terms of number and type of memory bus transactions are given in addition to area and power figures for the \sy hardware. While absolute and relative circuit complexity of the added hardware is comparable to our proposed solution, its power consumption of tens of milliwatt is in the range of our targeted total system consumption and, therefore, prohibitive \cite{Pullini19}. We follow the same approach of a memory-mapped shared peripheral with comparable circuit area but further reduce the cost of \sy primitives, state them in terms of cycles and energy and quantize the achieved system-wide energy savings while maintaining the overall power envelope of a few tens of milliwatts.
We consequently use a cycle-exact \gls{RTL} implementation of the whole system to measure execution time and a post-layout, fabrication-ready physical implementation in a current 22\,nm CMOS process as a basis for the most important figure of merit, the total system energy with and without our proposed solution for fine-grain parallelism. An important rationale for the post-layout implementation stage is that it considers the clock distribution network which typically consumes a significant share of the overall dynamic power of synchronous digital circuits, yet usually gets neglected in energy analyses. Furthermore, the efficacy of fine-grain \gls{PM} in the form of silencing parts of the clock network, a central part of our concept, can only be shown in this way.

\section{Background}\label{sec:background}

This work aims at accelerating \sy tasks in the context of \textit{tightly memory-coupled multicore processing clusters}. This type of system, such as Rigel \cite{Kelm09}, STHORM \cite{Benini12}, or PULP \cite{PULPv2}, is designed to cope with computation-intensive, data-centered processing tasks with a minimal amount of hardware complexity to allow the usage in very energy-efficient, yet performant \gls{IoT} end-nodes. The clusters are built around a set of RISC-type \glspl{PE} that have low-latency access to a shared \gls{SPM} (single cycle in our cluster variant). As a main consequence, this design does not require the adoption of area-intensive data caches, thus avoiding any coherency-ensuring overhead. Consequently, this type of cluster is also called \textit{cacheless shared-memory multiprocessor}.

Upper-bounding the core count to 16 in a single cluster allows the usage of rather simple, but fast (in terms of latency) interconnect and bus systems, while still providing computational performance of several GOPS. Other architectures like GP-GPUs scale the number of \glspl{PE} up to 32 with a two-cycle latency shared-memory, prioritizing performance over efficiency. This design choice stands in contrast to the target of this work, where we propose parallelism as a way to improve the energy-efficiency of a low-power computing system. Multiple instances of clusters, each connected to a higher level of shared memory, enable systems with performance demands that cannot be satisfied with a single cluster. In this way, the principles for \gls{NTC} that we outlined in \secref{sec:intro} are respected: As we confirm in \secref{sec:exp}, \glspl{PE} and memory consume the lion share of both dynamic and static power. As the performance of a single cluster already satisfies the requirements of several real-life applications (e.g., \cite{Palossi19,Kartsch19}), we set the focus of this work to single-cluster energy-efficient \sy and leave the hierarchical extension to multiple clusters for future work.

The \gls{SPM} must be realized with either multi-banked SRAMs or \gls{SCM} to enable \gls{PE}-concurrent and low-latency access to it. Furthermore, the low-latency constraint bounds the size of the \gls{SPM} to at most a few hundreds of kilobytes. Otherwise, the \gls{PE}-to-\gls{SPM} timing path, which defines the maximum operating frequency of a cluster, gets too long and results in degradation of the overall cluster performance. On the one hand, increasing the number of \gls{SPM} banks (i.e., scaling up memory size \textit{horizontally}) causes a drastic increase in interconnect latency \cite{Rahimi11}; on the other hand, scaling up memory \textit{vertically} while maintaining the number of banks increases the delay in address decoding for row selection.

\subsection{Relevance of fine-grain \sy support}\label{sec:bg:finegrain}

The aforementioned memory constraints limit the size of the working set that can be present in the tightly coupled memory at a given time. To avoid accesses to outer memory levels and preserve performance, applications with an extensive working set must employ techniques such as \emph{data tiling} to exploit data locality \cite{Tagliavini14}; a \gls{DMA} is additionally required to enable a double-buffering scheme (from/to a larger background memory with higher latency) and to overlap memory transfers with computation phases on the PEs. The orchestration of tiling introduces additional dimensions to the iteration space of the original algorithm, which map to supplementary inner loops iterating on smaller bounds (i.e., the tile size); consequently, \sy moves at a finer level of granularity than the original algorithm, and the number of \sy points increases by the number of tiles \cite{Tagliavini16}.

Another important aspect that contributes to the importance of fine-grain \sy support is to allow efficient parallelization of kernels that inherently exhibit small \glspl{SFR}. We list multiple examples of such kernels in \secref{sec:exp}, which can only be efficiently executed in a parallel fashion when the system supports the handling of typical \sy tasks in roughly ten cycles. The adoption of task-level parallelism, in combination with software pipelining, can work around these issues; however, this methodology poses a significant limitation in terms of programming flexibility and achievable computation latency. It furthermore requires the constant availability of tasks that can be independently scheduled and completely occupy the idle \glspl{PE}.

\begin{figure}[!t]
\centering{
  \includegraphics[width=1\columnwidth]{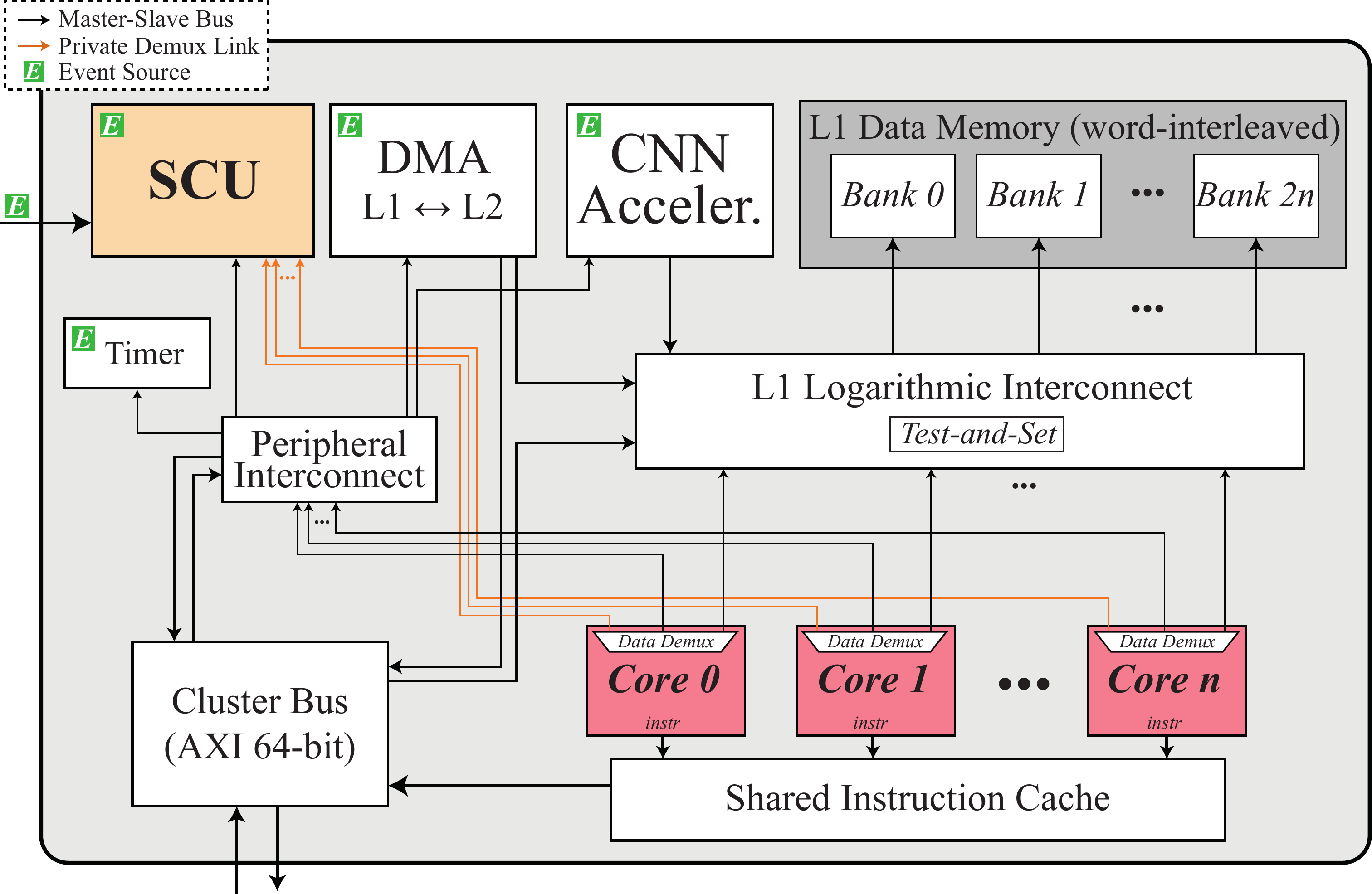}
  \caption{\label{fig:pulp_cluster}Multicore cluster, incorporating the proposed synchronization and communication unit (SCU).}
  }
\end{figure}
\subsection{Requirements for \sy hardware}

A beneficial adoption of the parallel \gls{NTC} concept requires adhering to a set of design principles for the cluster architecture. As the primary guideline, the overhead to provide \glspl{PE} with access to shared resources (such as memories or peripherals) and to communicate with each other must be kept as small as possible since a significant portion of the overall energy at the \gls{OEP} is spent through static power consumption, being directly linked to circuit area. Similarly, complex \sy hardware would cause a significant amount of extra active power that could eventually diminish any energy savings gained from accelerated computations.

As even simple interconnect systems or caches can quickly become comparable to or even exceed the area of the \glspl{PE} \cite{Pullini19}, consequently affecting the \gls{OEP} adversely, special care has to be taken when designing these building blocks and the clusters. As a result of this constraint, features such as multi-level data caches with the attached burden of coherency management, \glspl{MMU}, nested vectorized interrupt support, or network-like communication systems are unaffordable. The absence of these blocks, in turn, prohibits the usage in a control-centric OS-like fashion with virtual memory support but favors the employment of the clusters as \glspl{PMCA} to execute computation-centric kernels with regular program flow and physical memory addressing \cite{Vogel17}.

To not lose generality, the changes required in the host system should be minimal (e.g., no profound modifications or extensions to the \gls{PE} data path) and, wherever possible, existing infrastructure reused.
As the central \sy hardware is best aware of the set of \glspl{PE} that is waiting at \sy points, it is the best place to implement fine-grain \gls{PM} on a per-\gls{PE} basis. To consequently perform power-managing on any idle system component, one of the most important parallel \gls{NTC} principles, also the \sy hardware itself must be designed in an appropriate way to, e.g., not waste active power during phases were all \glspl{PE} are busy, and none is involved in any \sy action.

A solution that adheres to all stated requirements increases the energy efficiency of a cluster in two ways, where both essentially stem from reducing the execution time for a given task: First, the energy spent gets reduced proportionally to execution time as long as the power of the accelerated system is similar to that of the baseline system. Second, the reduction in execution time allows lowering the operating point (voltage and frequency) of the system, moving it closer to the \gls{OEP} for a given performance or latency target.

\begin{figure}[!t]
\centering{
  \includegraphics[width=1\columnwidth]{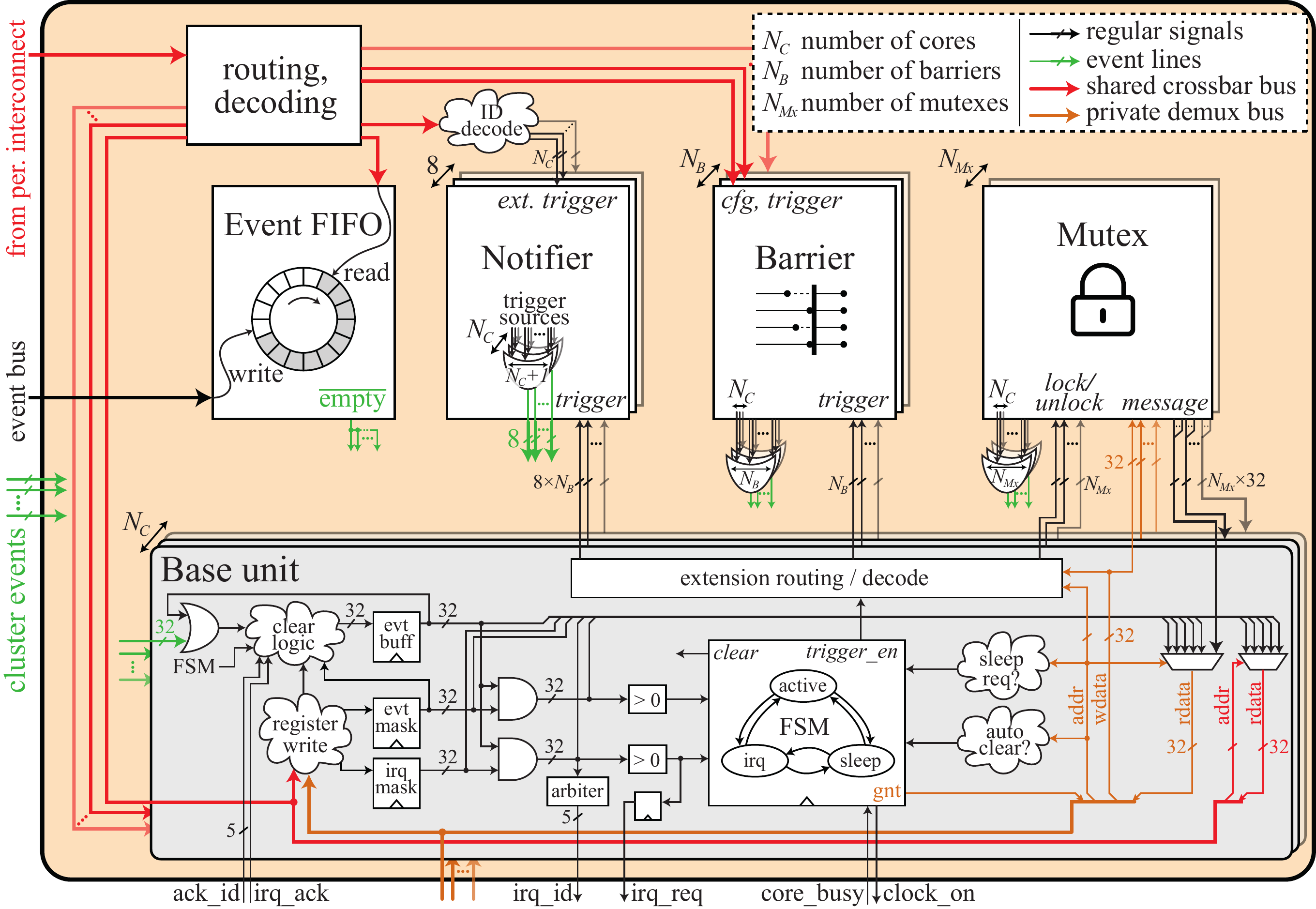}
  \caption{\label{fig:SCU_archi}Simplified overall architecture of the SCU, including the available extensions, and detailed architecture of the base units. Signals on the bottom connect to the cores, signals on the left to the cluster, and higher hierarchies.}
  }
\end{figure}
\section{Architecture}\label{sec:archi}
This section starts with an introduction to the high-level architecture of the hosting multiprocessor cluster before explaining the details of the \gls{SCU} architecture, its integration into the cluster as well as analyzing its scalability.
\subsection{Multiprocessor Cluster}\label{sec:archi_cluster}
As a basis for our proposed \sy solution, we use the open-source multiprocessor cluster of the PULP project \cite{Pullini19}, matching the targeted deeply embedded, data cache-less tightly-memory coupled system type. The cluster is designed around a configurable number (typically up to 16) of low-cost in-order RISC-V microcontrollers. To greatly accelerate the execution of the targeted \gls{DSP}-centric processing loads, they feature several extensions to the base instruction set \cite{Gautschi17} from which a wide range of applications benefit. Specialized \glspl{PE} such as neural network accelerators can additionally be included in the cluster to cope with more specific tasks that require very high processing throughput \cite{Conti17}.

All \glspl{PE} share a single-cycle accessible L1 \gls{TCDM}, composed of word-interleaved single-port SRAM macros. In order to reduce the number of contentions between \glspl{PE}, a banking factor of two is used (i.e., the number of banks is twice the number of \glspl{PE}). Data transfers between the size-limited \gls{TCDM} and the L2 memory with larger capacity is facilitated by a tightly-coupled \gls{DMA} connected to the L1 memory like any other \gls{PE}. Access routing and arbitration between all \glspl{PE} and the \gls{TCDM} banks are handled by a low-latency \gls{LINT}, allowing each \gls{TCDM} bank to be accessed by a \gls{PE} in every cycle. All the cores fetch the instructions from a hybrid private-shared instruction cache which has -- in addition to the \gls{DMA} -- access to the 64-bit AXI cluster bus that connects to the rest of the communication and memory system of the SoC.

In addition to the \gls{TCDM} interconnect, the cluster features a peripheral subsystem with dedicated \gls{LINT}. It allows not only the specialized \glspl{PE} to be programmed and controlled by means of memory-mapped configuration ports, but also to connect further peripherals such as timers and the \gls{SCU} that is proposed in this work. A master port to the cluster bus allows the RISC-V cores to access all cluster-external address space.

\subsubsection*{Test-and-Set Atomic Memory Access}
Besides routing and arbitrating requests, the \gls{LINT} provides basic and universal atomic memory access to the whole \gls{TCDM} address space in the form of \gls{TAS}. Atomic accesses are signaled by setting an address bit that is outside of the L1 address space; the \gls{LINT} checks it upon read-access. The currently stored value is returned to the requesting core (or to the elected one in case of multiple contending requests) and -1 written back to memory in the next cycle before any other core gets its request granted. Synchronization primitives based on this feature are used as a strong baseline (\gls{TAS} transactions take just three cycles) against which we compare our proposed solution.

\subsection{SCU Base Unit}
\figref{fig:SCU_archi} depicts a high-level overview of the \gls{SCU} architecture, with a deeper focus on the \gls{SCU} base unit.
The base unit is instantiated once per RISC-V core and provides the fundamental functionality of the \gls{SCU}, i.e., event and wait-state management, as well as fine-grain \gls{PM} through direct control of the clock-enable signal of the corresponding core. The design is based on 32 level-sensitive \textit{event lines} (per core) that are connected to associated \textit{event sources}. In a typical usage scenario, a limited number of the event sources are located outside of the \gls{SCU} (e.g., specialized \glspl{PE} or cluster-internal peripherals), while the remaining ones are responsible for core-to-core signaling and generated within the \gls{SCU} by so-called \gls{SCU} \textit{extensions}.

Event lines are stored into a register called \textit{event buffer}, which is maskable through the \textit{event mask} register.
Basic interrupt support is also provided to handle exceptions and other irregular events; an \textit{interrupt mask} register allows selective enabling and disabling of event lines to trigger hardware interrupts.
The central \gls{FSM} orchestrates all control flow and includes the three states, \textit{active, sleep}, and \textit{interrupt-handling}. The main inputs used to evaluate state transitions are pending events or interrupts, the core busy-status as well as sleep and buffer-clear requests.

\subsection{SCU Extensions}
\gls{SCU} \textit{extensions} are responsible for core-to-core signaling; generally, they generate core-specific events that allow a subset to continue execution. All extensions have trigger and configuration signals connected to each base unit; as with the base units, their associated functionality is available through memory-mapped addresses. The four available types of extensions are depicted in \figref{fig:SCU_archi} and detailed in the following.

\textbf{Notifier:} This extension provides general-purpose, any-to-any matrix-style core-to-core signaling. Each core can trigger one of the eight notifier events for any subset of cores (including itself). For write-triggered events, the write data are used as a target-core mask; for read-triggered events, a dedicated register in each \gls{SCU} base unit holds the target mask. An all-zero value causes a broadcast notifier to all cores in both cases. This extension is used in the \gls{TAS}-based variants of the \sy primitives that we profile and use in \secref{sec:synth_bench} and \secref{sec:exp_apps}, respectively.

\textbf{Barrier:} Allows a configurable \textit{target} subset of cores to continue execution only after a (possibly different) \textit{worker} subset has reached a specific point in the program. The extension contains a status register that keeps track of each core that has already arrived at the barrier; this is signaled by reading or writing from or to specific addresses. Depending on the core that caused the access, the matching bit in the status register is set. Once the status register matches the configured \textit{worker} subset, an event is generated for all cores that are activated in the \textit{target} subset, allowing those to uninterruptedly idle-wait at the barrier until their condition for continuation is met.

\textbf{Mutex:} Represents an object that can only be owned or locked by one core at a time and, therefore, directly supports \sy primitives that require mutual exclusivity such as, e.g., mutual exclusive code sections. Try-locks are, similar to the barrier extension, signaled by reading from a specific address. The mutex extension keeps track of all pending lock requests and elects one core by sending an event to only that core. The elected core must write to the same address once it releases the mutex, causing the extension to wake up another waiting core (if there is any).

\textbf{Event FIFO:} To be able to react to (relatively slow) cluster-external event sources as well, the event FIFO extension is included in the \gls{SCU}. It allows handling of up to 256 cluster-external event sources that can be triggered by, e.g., chip-level peripherals or higher-level control cores, as can be found in modern \glspl{SoC}. The external events are sequentially received over a simple request/grant asynchronous 8-bit event bus and stored into the FIFO. As long as there is at least one event present, an event line associated with the FIFO is asserted that is connected to all \gls{SCU} base units. In a typical use-case, the event line triggers an interrupt handler on one core that then pops the events from the FIFO and processes them.

For the targeted parallel programming models such as, e.g., OpenMP \cite{OpenMP}, the barrier and mutex extensions are the most important ones as they provide hardware support for the fundamental \textit{parallel sections} and \textit{critical sections} programming primitives.
The number of barrier and mutex extension instances, $N_B$ and $N_{Mx}$, can be independently set at design time to, e.g., support every team-building variant. As every core can only wait at one barrier or try-lock one mutex at a time, the corresponding core-specific events of all instances are combined into a single event per extension type and core.

\subsection{SCU Integration}
Being a part of the peripheral subsystem that is explained in \secref{sec:archi_cluster}, the \gls{SCU} is connected as an additional shared, memory-mapped peripheral to the corresponding \gls{LINT}. However, this single-port solution has the major drawbacks of non-deterministic core-to-\gls{SCU} access latency and sequentialized accesses whenever more than one core wants to access the \gls{SCU} in a given cycle. Since the limitations (in terms of performance, energy-efficiency, and scalability) of \sy primitives that are realized with classic atomic memory access largely result from sequential access to shared variables, parallel access to the \gls{SCU} base units and extensions responsible for core-to-core signaling is paramount.

We, therefore, use additional, private one-to-one buses between each core and its corresponding \gls{SCU} base unit, shown in orange in \figref{fig:pulp_cluster}, and \figref{fig:SCU_archi}. A demux at the data port of each core selects between the L1 \gls{TCDM} and peripheral \glspl{LINT} and the private \gls{SCU} link. The one-to-one correspondence between the cores and \gls{SCU} base units allows to alias their address space, thereby simplifying \sy primitives by removing core-id dependent address calculations. As the paths through the \glspl{LINT}, the private core-\gls{SCU} links are purely combinational and therefore allow for single-cycle access.
As we demonstrate in \secref{sec:synth_bench}, the fully-parallel access to the \gls{SCU} can even result in constant cycle cost for certain \sy primitives, independently from the number of involved cores -- a very favorable scaling property compared to classical atomic-memory based approaches.

In order to retain a global address space (for, e.g., debugging purposes), all base units are as well accessible from every core and from outside the cluster through the peripheral \gls{LINT}. All power-managing functionality of the \gls{SCU} base units (resulting in a core idle-waiting for an event) is not implemented for this access method as it would disturb the inter-core control flow.

\begin{figure}[!t]
\centering{
  \includegraphics[width=1\columnwidth]{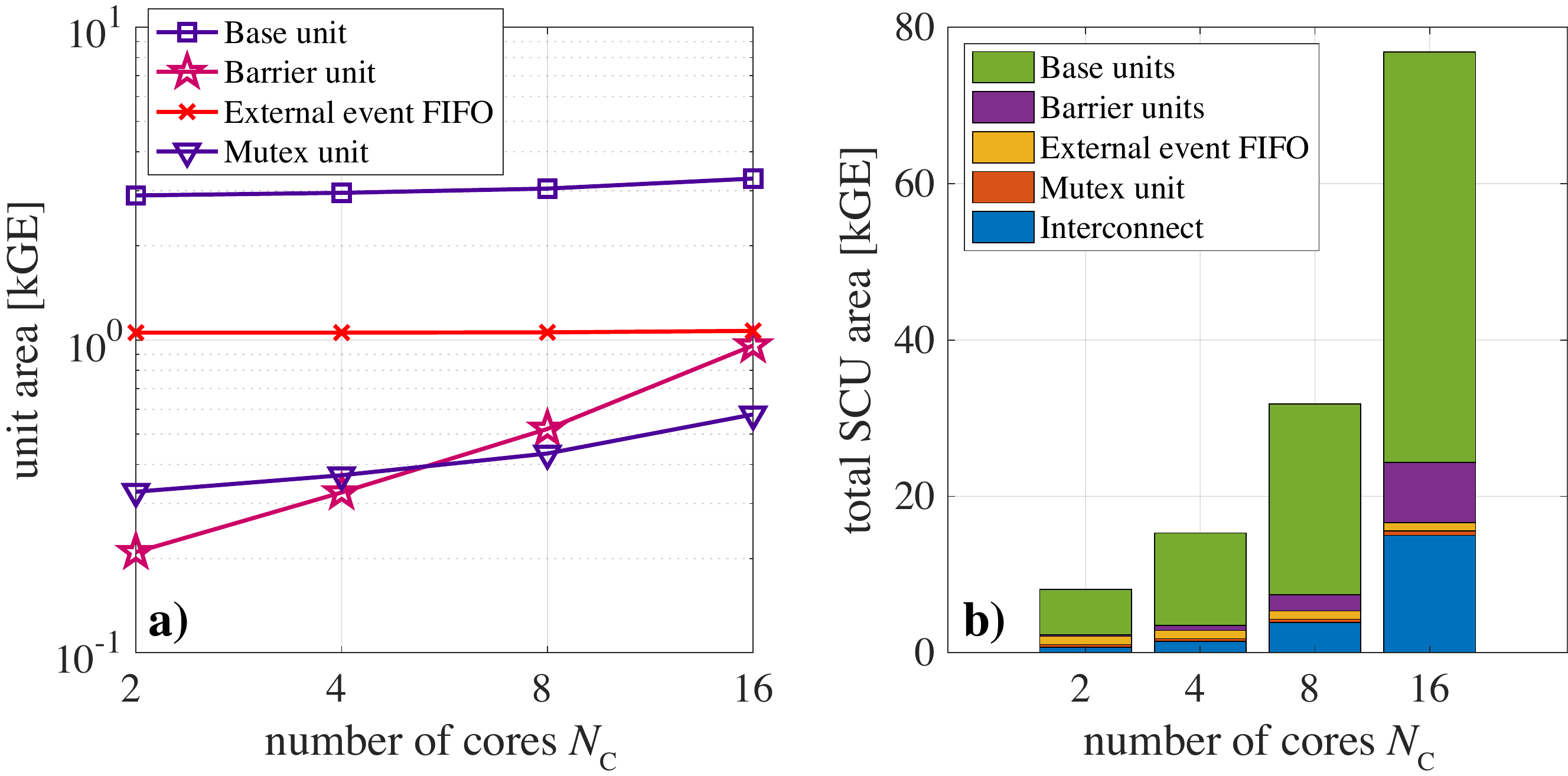}
  \caption{\label{fig:unit_area}Scaling of the circuit area (in gate-equivalents (GE)) for the base unit and the available extensions (a) and for the overall \gls{SCU} (b).}
  }
\end{figure}

\subsection{SCU Scalability}
\figref{fig:unit_area} shows both the total \gls{SCU} area as well as the area of the individual sub-units and extensions in relation to the number of cores $N_C$. For the total area, a typical configuration with the number of barrier extensions $N_B=N_C/2$, and the number of mutex extensions $N_{Mx}=1$, is shown. The plots show post-synthesis numbers; we used the same 22\,nm CMOS process as for the multicore cluster, which hosts the \gls{SCU}. Design synthesis was done in the slow-slow process corner, at 0.72\,V supply voltage, a temperature of 125\,$^\circ$C, and with a 500\,MHz timing constraint.\footnote{Even though our multicore clusters are usually constrained to slower clock frequencies (see \secref{sec:exp:method}), we chose this constraint to also verify the suitability of the \gls{SCU} for systems that target slightly higher clock speeds.} We restrict $N_C$ to a maximum of 16, matching the typical scalability limits of the targeted cluster-based architecture. An analysis of the slopes in the double-logarithmic sub-unit area plot in \figref{fig:unit_area}b) reveals a mildly super-linear scaling for the barrier extensions and sub-linear or constant scaling for the others. Overall \gls{SCU} area favorably scales sub-linearly up to the typically used configuration of $N_C=8$ and mildly super-linearly if $N_C$ is further increased to the maximum configuration. The area contribution of the \gls{SCU} base units and barrier extensions dominates in all configurations, however, the share of the \gls{SCU}-internal interconnect logic to correctly route all $N_C+1$ slave ports to the respectively connected sub-units becomes as well significant for the two largest configurations.

\section{Single-instruction Synchronization}\label{sec:single_sync}
With our goal of aggressively reducing \sy overhead in mind, we are proposing a scheme that allows handling common \sy tasks with the execution of a single instruction in each involved RISC-V core. To achieve this, we extensively leverage the dedicated link between each core and the corresponding \gls{SCU} base unit, the associated aliased address space of 1\,Kibit, and the possibility to stall a core by not granting accesses made over the private links.

\begin{figure}[!t]
\centering{
  \includegraphics[width=\columnwidth]{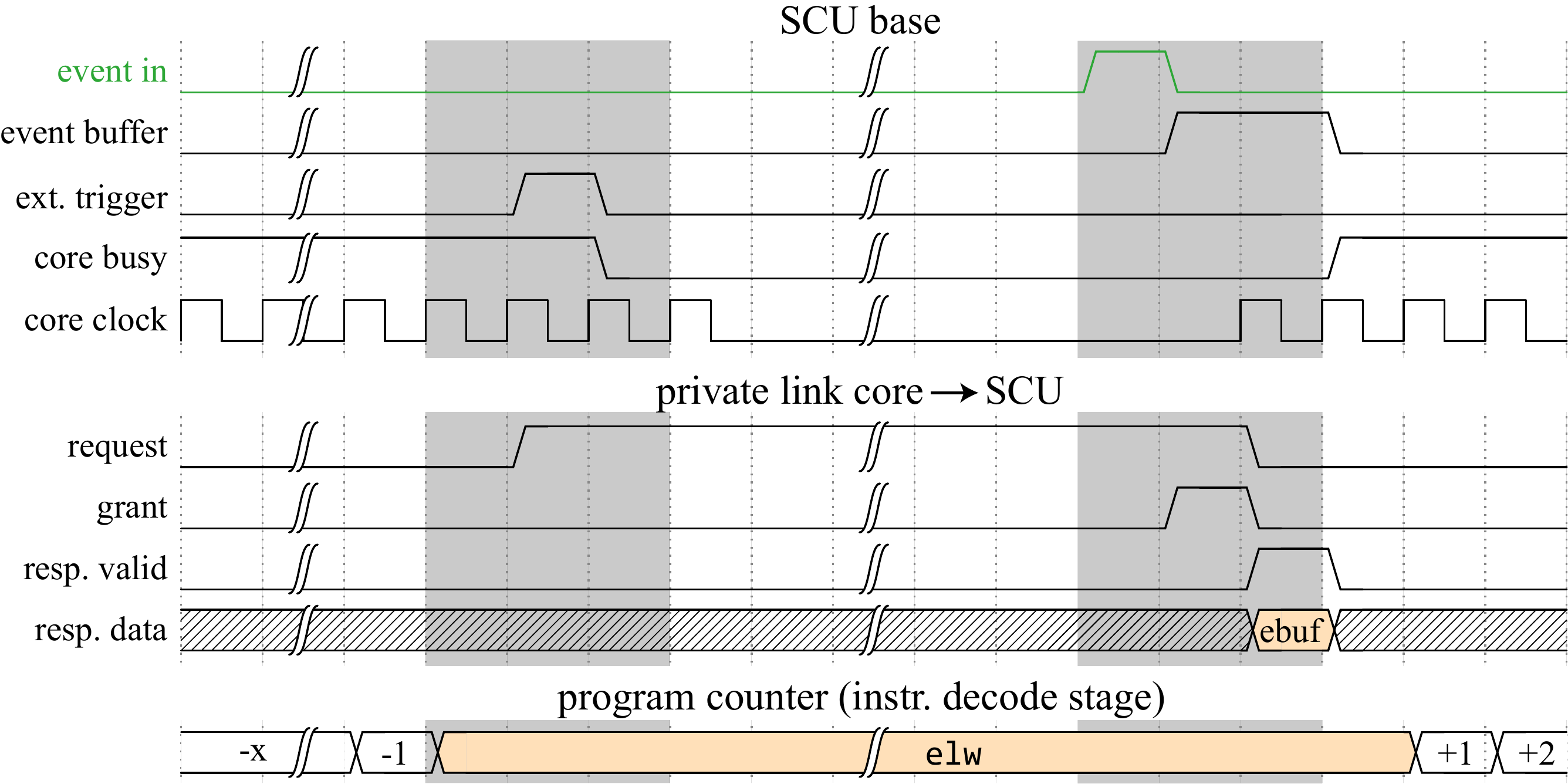}
  \caption{\label{fig:evt_sig}Interfacing of the SCU with a RISC-V core and corresponding timing. Shaded intervals correspond to transitions to (left) and from (right) sleep state, respectively.}
  }
\end{figure}

A fundamental aspect of our proposed solution is that whether a core can continue at a \sy point is always signaled through events that are generated inside the \gls{SCU} by one of the extensions. Each involved core idle-waits for the appropriate event to occur; the corresponding event line has to be activated in the event mask. Waiting is universally initiated by executing the \texttt{elw} instruction that we added to the extensible RISC-V \gls{ISA}. The mnemonic stands for \textit{event-load-word}; the instruction is identical to the regular \textit{load-word} (\texttt{lw}) of the base \gls{ISA} with the exception of an altered opcode such that the core controller can distinguish them. Whenever a core executes \texttt{elw} with an address that requests waiting for an event, the \gls{SCU} will block the resulting transaction on the private link by not asserting the grant signal (given that no events are currently registered in the event buffer). This process is depicted in the left shaded part of \figref{fig:evt_sig}, which shows the details of a private core-\gls{SCU} link and the most important signals of the corresponding \gls{SCU} base unit. Due to the in-order nature of the employed cores, the stall at the data port propagates through the core pipeline. The \texttt{elw} opcode causes the core controller to release the busy signal once any prior multi-cycle instructions have been executed; consecutively, the \gls{SCU} power-manages the requesting core by lowering its clock-enable signal. Depending on the address of the discussed read transaction, an extension in the \gls{SCU} gets simultaneously triggered (e.g., try-lock a mutex, set the status bit in a barrier, send a notifier event). The extension triggering is controlled by the \gls{FSM} in the \gls{SCU} base unit to ensure that per \texttt{elw}-transaction triggering happens only once.

The right shaded part of \figref{fig:evt_sig} shows the process of a core waking up and continuing execution, initiated by an incoming event. The event is present in the event buffer in the consecutive cycle, causing the \gls{SCU} both to re-enable the core clock and assert the grant for the still-pending read request. Another cycle later, the response channel of the private link is used to deliver additional information to the requesting core: Often, the content of the event buffer is sent such that in the case of multiple activated event lines, the core can immediately evaluate the reason for the return from sleep. More interestingly, however, the response channel can also be used to pass extension-specific data. In the example of the mutex extension, it allows the unlocking core, done with a write transaction, to intrinsically pass a 32-bit message to the core that locks the mutex next. Once the response data is consumed, the event buffer can -- again controlled through the address of the \texttt{elw} -- automatically be cleared, freeing cores from yet another common task, especially for the usual case of waiting for a single event line only.

\figref{fig:evt_sig} shows the process of entering and leaving wait state with an address that results in both triggering an extension and automatically clearing the buffer, additionally highlighting the small amount of only six cycles of active core clock for handling a \sy point (excluding the possibly required address calculation for \texttt{elw}). For cases where an active event occurs before or during a wait request (e.g., when the last core arrives at a barrier), the grant is immediately given, and no power-managing is done to not waste any cycles. The required changes in the core to support the described, powerful mechanism are limited to decoding the \texttt{elw} instruction to release the busy signal, which would otherwise remain asserted on a pipeline stall due to the pending load at the data port.

\subsection{Fused Interrupt Handling}
The targeted type of clusters is primarily meant for executing kernels with a regular program flow, the synchronization of which can be purely handled with events and idle waiting. Still, interrupts are often required to, e.g., handle data exceptions or react to other spontaneous, irregular, but important events that require an immediate change of program flow. A dedicated \gls{FSM} state and an additional mask register in each \gls{SCU} base unit are employed to fulfill said requirement; the event buffer is shared between both masks for increased area and energy efficiency. The few cases where a core needs to be sensitive to the same event source both as an interrupt and event trigger can be handled with a combination of an interrupt handler and a self-triggering notifier event. Two dedicated request/identifier pairs connect each core and the corresponding \gls{SCU} base for both requesting and clearing interrupts, respectively. The \gls{SCU} arbitrates one of the pending interrupts to the core, which, in turn, acknowledges the processing of interrupt identifiers upon entering the respective handler. Similar to the auto-clearing capability when waking up through events, the bit corresponding to the called interrupt handler is cleared in the event buffer to reduce management overhead in the handlers.

Should an active event occur during an interrupt handler, regular program flow is immediately continued after its termination. In the other, usual case, the \gls{FSM} transits to sleep again and awaits further incoming events and interrupts. After termination of the interrupt handler, the \texttt{elw} instruction responsible for the original wait state is re-executed, allowing the \gls{SCU} to detect said termination and power-manage the core again. In such cases, the \gls{FSM} takes care of inhibiting erroneous extension re-triggering upon the repeated sleep request after interrupt handling.

\section{Experimental Results}\label{sec:exp}
To demonstrate the effectiveness of the proposed event-based, hardware-supported \sy concept, we present two types of experimental results in this section. We first show the theoretically achievable improvements through synthetic benchmarks where all experiment parameters can be controlled. We successively analyze the performance and energy efficiency improvements that are observed when executing a range of applications that the targeted class of deeply embedded \glspl{CMP} is typically used for.
As the leading principle and motivation for this work is to reduce the energy that the cluster consumes for a given workload or task, we report not only the total cluster energy but also power and execution time in all cases to provide insight into how the energy reduction is achieved. We additionally provide power breakdowns into the main contributors to highlight the importance of fine-grain power management in the form of clock gating. Finally, an analysis of both the amount of total and active cycles spent on \sy shows how the proposed solution drastically reduces \sy-related overhead.

\subsection{Baseline}\label{sec:exp-baseline}
As a baseline, we use purely software-based implementations of \sy primitives that employ spin-locks on \gls{TAS}-protected variables in the L1 \gls{TCDM} with the help of the \gls{TAS}-feature of the logarithmic interconnect. Since most modern multiprocessor systems, however, feature hardware support for idle waiting, it is common to avoid thereby the active and continuous polling of \sy variables, which removes a very significant amount of core activity and memory accesses and therefore wasted energy as well as memory bandwidth.

To take account of this, we include a second \gls{TAS}-based solution in our comparisons where cores that do not succeed in acquiring a \sy variable (e.g., to update the barrier status stored in the variable) are put to sleep with the help of the \gls{SCU} (by idle waiting on an event as described in \secref{sec:single_sync}). Whenever the current owner updates or releases the variable, it also uses an \gls{SCU} notifier broadcast event to wake up all the remaining, sleeping cores, which will then again try to acquire the \sy variable. While the described \sy mechanism can also be realized with similar solutions for idle waiting and notifying cores that may be found in comparable systems, it -- in our case -- already benefits from the low latencies for notifiers and idle state handling that are enabled through the \gls{SCU}.
In the following, the purely spin-lock based implementations of \sy primitives will be referred to as \textit{SW} and the idle-waiting extended versions as \textit{TAS}.

\subsection{Experimental Setup and Methodology}\label{sec:exp:method}

All experiments were carried out on an eight-core implementation of the multicore cluster of \figref{fig:pulp_cluster}. It features 64\,kByte L1 \gls{TCDM} and eight kByte of shared instruction cache; the \gls{SCU} contains four barrier and one mutex extensions. For all cycle-based results, an \gls{RTL} description of the cluster and cycle-exact simulations were used. In order to obtain more detailed insight than total execution time only, a range of non-synthesizable observation tools both in the \gls{RTL} description of the RISC-V cores as well as in the testbench is employed. Per-core performance counters record the number of executed instructions, stalls at both data and instruction ports, and the like; an instruction tracer allows detailed analysis of the executed applications and benchmarks. Overall execution time is measured with the help of a cluster-global timer that is part of the cluster peripherals and also present in the fabricated \gls{ASIC}. The timer is activated only during periods where the actual benchmark (synthetic or application) is executed to exclude, e.g., initialization and boot periods. The enable signal of the timer is monitored in the testbench and the timestamps of rising and falling edges recorded, allowing to identify the relevant portions of the trace files.

\subsubsection*{Physical implementation}
To obtain power (and ultimately energy) results, the \gls{RTL} model of the cluster was synthesized in a 22\,nm CMOS process and a placed and routed physical implementation created; both steps were done with a 350\,MHz timing constraint and in the slow-slow process corner at 0.72\,V supply voltage and a temperature of 125\,$^\circ$C\footnote{Timing was verified with all permutations of the slow-slow/fast-fast process corners, 0.72\,V/0.88\,V supply voltage, and temperatures of -40\,$^\circ$C/125\,$^\circ$C.}. The resulting fabrication-ready and functionally verified module\footnote{Multiple \glspl{ASIC} containing very similar clusters are silicon proven in various technology nodes \cite{Flamand18,Pullini19,Schoenle18}.} measures 0.8\,mm$\times$1.4\,mm with pre-placed SRAM macros for the L1 \gls{TCDM}; the \gls{SCU} accounts for less than 2\% of the total circuit area. Both the synthetic benchmarks and the applications were run on the resulting gate-level netlist and activity files recorded during the benchmarking periods for every electrical net in the cluster. Again, the enable signal of the cluster-global timer is used to start and stop the activity recording. The subsequent hierarchical power analysis was done in the typical-typical process corner at 0.8\,V supply voltage and 25\,$^\circ$C, allowing us to report not only the total power and energy but also the respective breakdowns to analyze the contributions of the individual cluster building blocks. The reported power and energy results correspond to a cluster operating frequency of 350\,MHz.

\subsection{Synthetic Benchmarks}\label{sec:synth_bench}
We start our analysis by quantifying the cost in terms of cycles and energy for executing barriers and critical sections, the two \sy primitives that are most commonly used in the targeted parallel programming models. We compare the hardware variants featured in the \gls{SCU} with the purely spin-lock based as well as with the idle-wait extended baseline variants as described in \secref{sec:exp-baseline}. To highlight the favorable scaling behavior (with respect to the number of participating cores) of the \gls{SCU}, we provide the quantification for two, four, and eight cores even though the cluster is mainly designed for execution on all eight cores. To remove any edge effects such as instruction cache misses from the results, we let the involved cores execute a loop eight times that contains the respective primitive 32 times and average the resulting cycle count. A typical use-case for critical sections in the type of targeted systems is the placement at the end of an \gls{SFR} to perform small control tasks like updating a shared variable by all worker cores. Consequently, the critical section is usually very short (up to ten cycles only), a circumstance that we consider in our experiments.

The synthetic benchmarks were compiled with an extended version of the riscv-gcc 7.1.1 toolchain that supports the \texttt{elw} instruction, using the \texttt{-O3} flag. To compute not only the absolute cost figures that are reported in \tblref{tbl:primitive_cost} but also the relative energy overhead, we additionally measure the power during the execution of 512 \texttt{nop} instructions on a varying number of cores. While the choice of the \texttt{nop} instruction may intuitively not be a suitable representation of actual processing loads, the resulting relation between \gls{SFR} size and overhead still is a very reasonable estimate for the behavior that results when executing actual applications, as our analysis in \secref{sec:exp_apps} shows.

\begin{table}[!b]
\begin{center}
\caption{Cost of \sy primitives in terms of cycles and energy.}\label{tbl:primitive_cost}
\setlength{\tabcolsep}{2pt}
\begin{tabular}{p{55pt}C{30pt}R{16pt}R{16pt}R{16pt}C{4pt}R{16pt}R{16pt}R{16pt}}
                                &                     & \multicolumn{3}{c}{\textit{cycles}}                                    & & \multicolumn{3}{c}{\textit{energy [nJ]}}                                         \\
\cline{3-5}\cline{7-9}
$N_C$ \textit{(core count)}             &                     & \ctab 2               & \ctab 4               & \ctab 8                & & \ctab 2                  & \ctab 4                  & \ctab 8                    \\
\toprule
\multirow{3}{*}{Barrier}        & SCU                 & \cycBarrSCUcoreII     & \cycBarrSCUcoreIV     & \cycBarrSCUcoreVIII    & & \energyBarrSCUcoreII     & \energyBarrSCUcoreIV     & \energyBarrSCUcoreVIII     \\
                                & TAS                 & \cycBarrTAScoreII     & \cycBarrTAScoreIV     & \cycBarrTAScoreVIII    & & \energyBarrTAScoreII     & \energyBarrTAScoreIV     & \energyBarrTAScoreVIII     \\
                                & SW                  & \cycBarrSWcoreII      & \cycBarrSWcoreIV      & \cycBarrSWcoreVIII     & & \energyBarrSWcoreII      & \energyBarrSWcoreIV      & \energyBarrSWcoreVIII      \\
\midrule
\multirow{3}{*}{5-cycle crit. sect.}  & SCU           & \cycMutexVSCUcoreII  & \cycMutexVSCUcoreIV  & \cycMutexVSCUcoreVIII & & \energyMutexVSCUcoreII  & \energyMutexVSCUcoreIV  & \energyMutexVSCUcoreVIII  \\
                                & TAS                 & \cycMutexVTAScoreII  & \cycMutexVTAScoreIV  & \cycMutexVTAScoreVIII & & \energyMutexVTAScoreII  & \energyMutexVTAScoreIV  & \energyMutexVTAScoreVIII  \\
                                & SW                  & \cycMutexVSWcoreII   & \cycMutexVSWcoreIV   & \cycMutexVSWcoreVIII  & & \energyMutexVSWcoreII   & \energyMutexVSWcoreIV   & \energyMutexVSWcoreVIII   \\
\midrule
\multirow{3}{*}{10-cycle crit. sect.} & SCU           & \cycMutexXSCUcoreII   & \cycMutexXSCUcoreIV   & \cycMutexXSCUcoreVIII  & & \energyMutexXSCUcoreII   & \energyMutexXSCUcoreIV   & \energyMutexXSCUcoreVIII   \\
                                & TAS                 & \cycMutexXTAScoreII   & \cycMutexXTAScoreIV   & \cycMutexXTAScoreVIII  & & \energyMutexXTAScoreII   & \energyMutexXTAScoreIV   & \energyMutexXTAScoreVIII   \\
                                & SW                  & \cycMutexXSWcoreII    & \cycMutexXSWcoreIV    & \cycMutexXSWcoreVIII   & & \energyMutexXSWcoreII    & \energyMutexXSWcoreIV    & \energyMutexXSWcoreVIII    \\
\bottomrule
\end{tabular}
\end{center}
\end{table}

\begin{figure*}[!ht]
\centering{
  \includegraphics[width=1\textwidth]{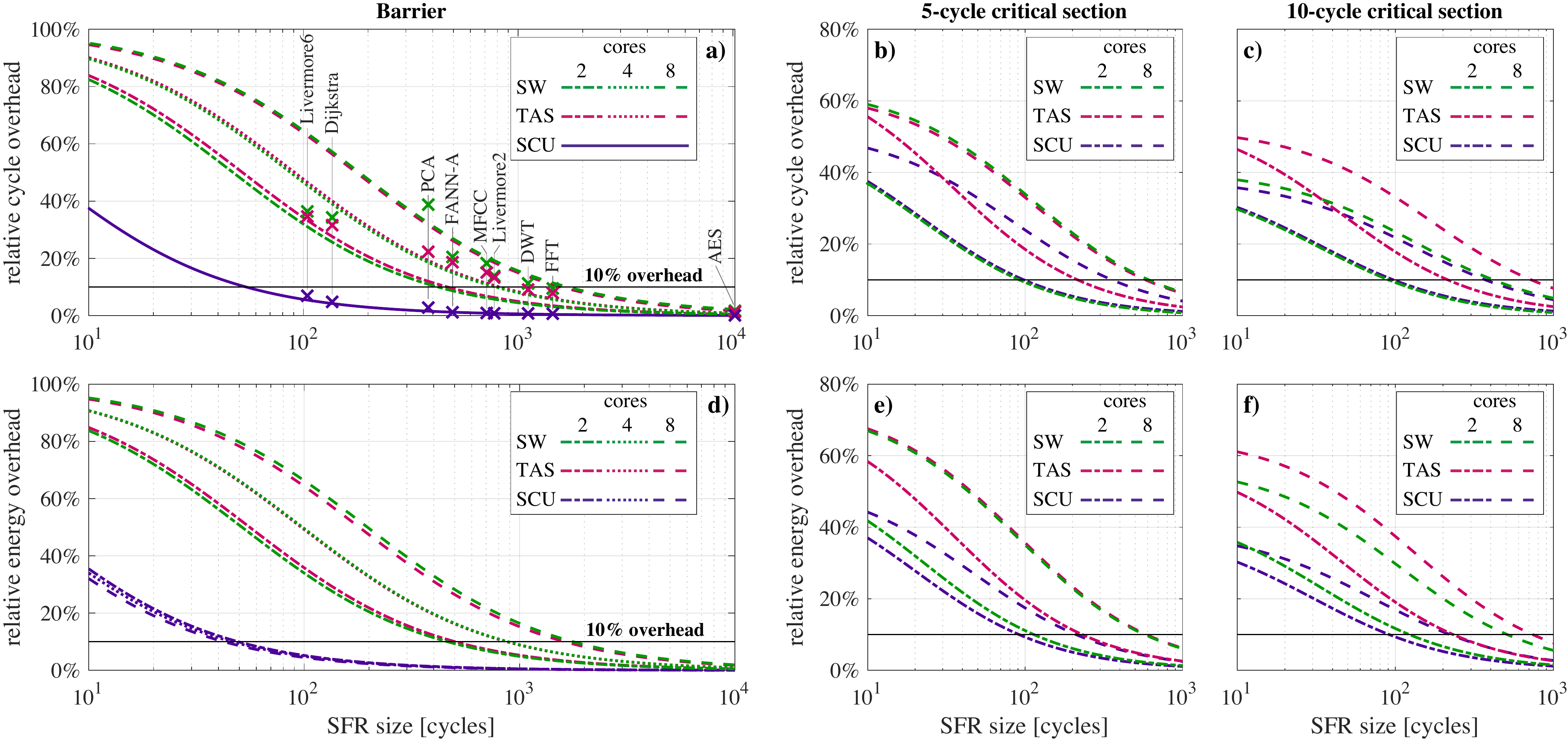}
  \caption{\label{fig:results_synthetic}Relative overhead in terms of cycles (a)-c), top) and energy (d)-f), bottom) vs. \gls{SFR} size for the three realizations of barriers (a), d), left) and critical sections with two lengths (b), c), e), f), right). The markers in a) indicate the relative share of active \sy cycles for the range of \gls{DSP} applications discussed in \secref{sec:exp_apps}. For critical sections, the graph lines corresponding to four cores have been omitted to improve graph readability. The scaling behavior in terms of overhead vs. core count is strictly monotonic (see raw costs in \tblref{tbl:primitive_cost}).}
  }
\end{figure*}

\subsubsection*{Barriers}
When considering the pure primitive cost, the \gls{SCU} barrier requires between \edgeSFRcycleBarrRatioSWcoreII$\times$ (2 cores, SW) and \edgeSFRcycleBarrRatioTAScoreVIII$\times$ (8 cores, SW and \gls{TAS}) fewer cycles, as can be seen in \tblref{tbl:primitive_cost}. The gap widens when considering energy where the reduction ranges between \energyBarrRatioTAScoreII$\times$ (2 cores, SW and \gls{TAS}) and \energyBarrRatioTAScoreVIII$\times$ (8 cores, TAS) or \energyBarrRatioSWcoreVIII$\times$ (8 cores, SW), respectively. With higher core counts, the \gls{TAS} version shows slightly lower energy compared to the SW version thanks to the idle-wait behavior. For the \gls{SCU} variant, not surprisingly, the fully parallel access to the barrier extension makes the cycle cost independent from the number of cores and incurs very little additional energy when increasing the number of participants. As a result, the \gls{SCU} supported barrier is especially favorable when a task is parallelized on all eight cores, which is the intended way of using the cluster.

\figref{fig:results_synthetic}a) and d) illustrate the raw barrier cost in relation to a preceding \gls{SFR} of varying size by showing the relative overhead for executing the barrier in terms of cycles and energy. While significant overhead reductions can be observed with \glspl{SFR} of up to around 1000\xspace cycles and eight active cores, the graphs reveal another even more important characteristic of the \gls{SCU} barrier: With a typical constraint of allowing up to 10\% of \sy overhead, the \gls{SCU} drastically reduces the smallest allowable \gls{SFR}. The cycle-related relative minimum \gls{SFR} reductions are (mathematically) identical to those for the raw primitive-cost; the energy-related reductions show only insignificant differences compared to the corresponding raw cost ratios. Besides the relative overhead reductions, the absolute size of the smallest allowable \gls{SFR} is important, which is with the \gls{SCU} barrier for both cycle and energy overhead and all core counts below 100\xspace cycles and therefore matches the in \secref{sec:relwork} stated requirement for fine-grain \sy. This is in stark contrast to the overheads resulting from the \gls{TAS} and SW variants, where both cycle and energy-based \gls{SFR}sizes must be at least multiple hundreds of cycles when considering two or four participating cores. The energy-related minimum \gls{SFR} with all eight cores participating, representing the most important case, is with \edgeSFRenergyBarrTAScoreVIII\xspace cycles (\gls{TAS}) and \edgeSFRenergyBarrSWcoreVIII\xspace cycles (SW) two orders of magnitude higher than the corresponding \gls{SFR} size when employing the \gls{SCU} barrier (\edgeSFRenergyBarrSCUcoreVIII\xspace cycles) and poses a strong limitation on the range of applications that can be efficiently parallelized on the targeted architecture.

\subsubsection*{Critical Sections}
Compared to barriers, the critical or mutual exclusive section \sy primitive can be more easily implemented with basic atomic memory access. The ability to enter the critical section can be managed with a single \gls{TAS}-protected variable that needs to be tested upon entering and written with the test value by the owning core upon exiting. For the \gls{TAS}-variant of this primitive, we link each access to the \sy variable to the usage of a notifier event to avoid constant testing of the variable by all cores that are waiting to enter the critical section: Any core that fails to enter will idle-wait for said event. The core that currently executes the critical section triggers the event upon exiting, causing all queued cores to quickly wake up and re-test the \gls{TAS}-variable, with all but the elected one immediately going back to sleep afterward.

In the \gls{SCU}-based implementation, we simply execute \texttt{elw} with an address mapped to the mutex extension, which elects one core for which continuation is enabled through the generation of a core-specific event. All others idle-wait at the mutex load until they are elected. Similar to the variants based on \gls{TAS}-variables, a write to the mutex by the previously elected core upon leaving the critical section unlocks the mutex and triggers the election of the next core to enter the section alongside the appropriate event generation. The distinction between locking and unlocking the mutex is done with the access type (read/write) and allows to use the same address for both operations, further reducing the software overhead for the \sy primitive.

As with barriers, we provide both raw-primitive cost (\tblref{tbl:primitive_cost}) as well as relative overheads in terms of performance and energy (\figref{fig:results_synthetic}), each, for two different critical section sizes. The latter is necessary since the wait behavior of cores that yet have to enter the critical section greatly differs between the implementations: For the SW variant, waiting cores do not only test the \sy variable upon another one exiting the critical section but constantly. Consequently, the duration of the critical section has an impact on so-caused parasitic energy. We calculate the overhead as the difference between ideal cycle count and energy and the measured ones. The ideal number of cycles is $T_{ideal}=N_CT_{crit}$ and the ideal energy $E_{ideal}=T_{ideal}P_{comp,1}$ with $T_{crit}$ denoting the length of the critical section, $P_{comp,1}$ the single-core cluster power and $N_C$ the number of cores that need to execute the critical section.

\begin{table*}[!t]
\caption{Main properties and results for the range of benchmarked \gls{DSP} applications. Active cycles reflect core-active cycles, i.e., cycles where the core clock is active, averaged over all eight cores.}\label{tbl:app_numbers}
\setlength{\tabcolsep}{1pt}
\begin{center}
\begin{tabular}{L{45pt}L{70pt}C{30pt}C{25pt}C{8pt}R{32pt}C{8pt}R{22pt}C{12pt}rC{5pt}rrC{15pt}rrC{5pt}rrC{10pt}c}
\textbf{Name}                 & \textbf{Domain}                 & \multicolumn{2}{c}{\textbf{Barrier}}                 &   & \ctab\textbf{SFR Size}  & & \ctab\textbf{Energy}      & & \multicolumn{4}{c}{\textbf{Execution Cycles}}                          & & \multicolumn{5}{c}{\textbf{Synchronization Cycles}}                                                                  & & \textbf{IPC} \\
\cline{3-4}\cline{10-13}\cline{15-19}
                              &                                 & \textit{count}                       & \textit{type} &   & \ctab\textit{[cycles]}  & & \ctab\textit{[\textmu J]} & & \ctab\textit{total} & & \multicolumn{2}{c}{\textit{active (stddev)}}   & & \multicolumn{2}{c}{\textit{total}}                 & & \multicolumn{2}{c}{\textit{active}}                 & &             \\
\toprule
\multirow{3}{*}{\appIname}    & \multirow{3}{*}{\appIdomain}    & \multirow{3}{*}{\appInumBarr}        & SCU           &   & \appISFRsizeSCU         & & \appIenergySCU       & & \appIexCycTotSCU    & & \appIexCycActSCU    & (\appIexCycActSprSCU   ) & & \appIsyncCycTotSCU    & (\appIsyncCycTotRelSCU   ) & & \appIsyncCycActSCU     & (\appIsyncCycActRelSCU   ) & & \appIIPCSCU \\
                              &                                 &                                      & TAS           &   & \appISFRsizeTAS         & & \appIenergyTAS       & & \appIexCycTotTAS    & & \appIexCycActTAS    & (\appIexCycActSprTAS   ) & & \appIsyncCycTotTAS    & (\appIsyncCycTotRelTAS   ) & & \appIsyncCycActTAS     & (\appIsyncCycActRelTAS   ) & & \appIIPCTAS \\
                              &                                 &                                      & SW            &   & \appISFRsizeSW          & & \appIenergySW        & & \appIexCycTotSW     & & \appIexCycActSW     & (\appIexCycActSprSW    ) & & \appIsyncCycTotSW     & (\appIsyncCycTotRelSW    ) & & \appIsyncCycActSW      & (\appIsyncCycActRelSW    ) & & \appIIPCSW  \\
\midrule
\multirow{3}{*}{\appIIname}   & \multirow{3}{*}{\appIIdomain}   & \multirow{3}{*}{\appIInumBarr}       & SCU           &   & \appIISFRsizeSCU        & & \appIIenergySCU      & & \appIIexCycTotSCU   & & \appIIexCycActSCU   & (\appIIexCycActSprSCU  ) & & \appIIsyncCycTotSCU   & (\appIIsyncCycTotRelSCU  ) & & \appIIsyncCycActSCU    & (\appIIsyncCycActRelSCU  ) & & \appIIIPCSCU \\
                              &                                 &                                      & TAS           &   & \appIISFRsizeTAS        & & \appIIenergyTAS      & & \appIIexCycTotTAS   & & \appIIexCycActTAS   & (\appIIexCycActSprTAS  ) & & \appIIsyncCycTotTAS   & (\appIIsyncCycTotRelTAS  ) & & \appIIsyncCycActTAS    & (\appIIsyncCycActRelTAS  ) & & \appIIIPCTAS \\
                              &                                 &                                      & SW            &   & \appIISFRsizeSW         & & \appIIenergySW       & & \appIIexCycTotSW    & & \appIIexCycActSW    & (\appIIexCycActSprSW   ) & & \appIIsyncCycTotSW    & (\appIIsyncCycTotRelSW   ) & & \appIIsyncCycActSW     & (\appIIsyncCycActRelSW   ) & & \appIIIPCSW  \\
\midrule
\multirow{3}{*}{\appIIIname}  & \multirow{3}{*}{\appIIIdomain}  & \multirow{3}{*}{\appIIInumBarr}      & SCU           &   & \appIIISFRsizeSCU       & & \appIIIenergySCU     & & \appIIIexCycTotSCU  & & \appIIIexCycActSCU  & (\appIIIexCycActSprSCU ) & & \appIIIsyncCycTotSCU  & (\appIIIsyncCycTotRelSCU ) & & \appIIIsyncCycActSCU   & (\appIIIsyncCycActRelSCU ) & & \appIIIIPCSCU \\
                              &                                 &                                      & TAS           &   & \appIIISFRsizeTAS       & & \appIIIenergyTAS     & & \appIIIexCycTotTAS  & & \appIIIexCycActTAS  & (\appIIIexCycActSprTAS ) & & \appIIIsyncCycTotTAS  & (\appIIIsyncCycTotRelTAS ) & & \appIIIsyncCycActTAS   & (\appIIIsyncCycActRelTAS ) & & \appIIIIPCTAS \\
                              &                                 &                                      & SW            &   & \appIIISFRsizeSW        & & \appIIIenergySW      & & \appIIIexCycTotSW   & & \appIIIexCycActSW   & (\appIIIexCycActSprSW  ) & & \appIIIsyncCycTotSW   & (\appIIIsyncCycTotRelSW  ) & & \appIIIsyncCycActSW    & (\appIIIsyncCycActRelSW  ) & & \appIIIIPCSW  \\
\midrule
\multirow{3}{*}{\appIVname}   & \multirow{3}{*}{\appIVdomain}   & \multirow{3}{*}{\appIVnumBarr}       & SCU           &   & \appIVSFRsizeSCU        & & \appIVenergySCU      & & \appIVexCycTotSCU   & & \appIVexCycActSCU   & (\appIVexCycActSprSCU  ) & & \appIVsyncCycTotSCU   & (\appIVsyncCycTotRelSCU  ) & & \appIVsyncCycActSCU    & (\appIVsyncCycActRelSCU  ) & & \appIVIPCSCU \\
                              &                                 &                                      & TAS           &   & \appIVSFRsizeTAS        & & \appIVenergyTAS      & & \appIVexCycTotTAS   & & \appIVexCycActTAS   & (\appIVexCycActSprTAS  ) & & \appIVsyncCycTotTAS   & (\appIVsyncCycTotRelTAS  ) & & \appIVsyncCycActTAS    & (\appIVsyncCycActRelTAS  ) & & \appIVIPCTAS \\
                              &                                 &                                      & SW            &   & \appIVSFRsizeSW         & & \appIVenergySW       & & \appIVexCycTotSW    & & \appIVexCycActSW    & (\appIVexCycActSprSW   ) & & \appIVsyncCycTotSW    & (\appIVsyncCycTotRelSW   ) & & \appIVsyncCycActSW     & (\appIVsyncCycActRelSW   ) & & \appIVIPCSW  \\
\midrule
\multirow{3}{*}{\appVname}    & \multirow{3}{*}{\appVdomain}    & \multirow{3}{*}{\appVnumBarr}        & SCU           &   & \appVSFRsizeSCU         & & \appVenergySCU       & & \appVexCycTotSCU    & & \appVexCycActSCU    & (\appVexCycActSprSCU   ) & & \appVsyncCycTotSCU    & (\appVsyncCycTotRelSCU   ) & & \appVsyncCycActSCU     & (\appVsyncCycActRelSCU   ) & & \appVIPCSCU \\
                              &                                 &                                      & TAS           &   & \appVSFRsizeTAS         & & \appVenergyTAS       & & \appVexCycTotTAS    & & \appVexCycActTAS    & (\appVexCycActSprTAS   ) & & \appVsyncCycTotTAS    & (\appVsyncCycTotRelTAS   ) & & \appVsyncCycActTAS     & (\appVsyncCycActRelTAS   ) & & \appVIPCTAS \\
                              &                                 &                                      & SW            &   & \appVSFRsizeSW          & & \appVenergySW        & & \appVexCycTotSW     & & \appVexCycActSW     & (\appVexCycActSprSW    ) & & \appVsyncCycTotSW     & (\appVsyncCycTotRelSW    ) & & \appVsyncCycActSW      & (\appVsyncCycActRelSW    ) & & \appVIPCSW  \\
\midrule
\multirow{3}{*}{\appVIname}   & \multirow{3}{*}{\appVIdomain}   & \multirow{3}{*}{\appVInumBarr}       & SCU           &   & \appVISFRsizeSCU        & & \appVIenergySCU      & & \appVIexCycTotSCU   & & \appVIexCycActSCU   & (\appVIexCycActSprSCU  ) & & \appVIsyncCycTotSCU   & (\appVIsyncCycTotRelSCU  ) & & \appVIsyncCycActSCU    & (\appVIsyncCycActRelSCU  ) & & \appVIIPCSCU \\
                              &                                 &                                      & TAS           &   & \appVISFRsizeTAS        & & \appVIenergyTAS      & & \appVIexCycTotTAS   & & \appVIexCycActTAS   & (\appVIexCycActSprTAS  ) & & \appVIsyncCycTotTAS   & (\appVIsyncCycTotRelTAS  ) & & \appVIsyncCycActTAS    & (\appVIsyncCycActRelTAS  ) & & \appVIIPCTAS \\
                              &                                 &                                      & SW            &   & \appVISFRsizeSW         & & \appVIenergySW       & & \appVIexCycTotSW    & & \appVIexCycActSW    & (\appVIexCycActSprSW   ) & & \appVIsyncCycTotSW    & (\appVIsyncCycTotRelSW   ) & & \appVIsyncCycActSW     & (\appVIsyncCycActRelSW   ) & & \appVIIPCSW  \\
\midrule
\multirow{3}{*}{\appVIIname}  & \multirow{3}{*}{\appVIIdomain}  & \multirow{3}{*}{\appVIInumBarr}      & SCU           &   & \appVIISFRsizeSCU       & & \appVIIenergySCU     & & \appVIIexCycTotSCU  & & \appVIIexCycActSCU  & (\appVIIexCycActSprSCU ) & & \appVIIsyncCycTotSCU  & (\appVIIsyncCycTotRelSCU ) & & \appVIIsyncCycActSCU   & (\appVIIsyncCycActRelSCU ) & & \appVIIIPCSCU \\
                              &                                 &                                      & TAS           &   & \appVIISFRsizeTAS       & & \appVIIenergyTAS     & & \appVIIexCycTotTAS  & & \appVIIexCycActTAS  & (\appVIIexCycActSprTAS ) & & \appVIIsyncCycTotTAS  & (\appVIIsyncCycTotRelTAS ) & & \appVIIsyncCycActTAS   & (\appVIIsyncCycActRelTAS ) & & \appVIIIPCTAS \\
                              &                                 &                                      & SW            &   & \appVIISFRsizeSW        & & \appVIIenergySW      & & \appVIIexCycTotSW   & & \appVIIexCycActSW   & (\appVIIexCycActSprSW  ) & & \appVIIsyncCycTotSW   & (\appVIIsyncCycTotRelSW  ) & & \appVIIsyncCycActSW    & (\appVIIsyncCycActRelSW  ) & & \appVIIIPCSW  \\
\midrule
\multirow{3}{*}{\appVIIIname} & \multirow{3}{*}{\appVIIIdomain} & \multirow{3}{*}{\appVIIInumBarr}     & SCU           &   & \appVIIISFRsizeSCU      & & \appVIIIenergySCU    & & \appVIIIexCycTotSCU & & \appVIIIexCycActSCU & (\appVIIIexCycActSprSCU) & & \appVIIIsyncCycTotSCU & (\appVIIIsyncCycTotRelSCU) & & \appVIIIsyncCycActSCU  & (\appVIIIsyncCycActRelSCU) & & \appVIIIIPCSCU \\
                              &                                 &                                      & TAS           &   & \appVIIISFRsizeTAS      & & \appVIIIenergyTAS    & & \appVIIIexCycTotTAS & & \appVIIIexCycActTAS & (\appVIIIexCycActSprTAS) & & \appVIIIsyncCycTotTAS & (\appVIIIsyncCycTotRelTAS) & & \appVIIIsyncCycActTAS  & (\appVIIIsyncCycActRelTAS) & & \appVIIIIPCTAS \\
                              &                                 &                                      & SW            &   & \appVIIISFRsizeSW       & & \appVIIIenergySW     & & \appVIIIexCycTotSW  & & \appVIIIexCycActSW  & (\appVIIIexCycActSprSW ) & & \appVIIIsyncCycTotSW  & (\appVIIIsyncCycTotRelSW ) & & \appVIIIsyncCycActSW   & (\appVIIIsyncCycActRelSW ) & & \appVIIIIPCSW  \\
\midrule
\multirow{3}{*}{\appIXname}   & \multirow{3}{*}{\appIXdomain}   & \multirow{3}{*}{\appIXnumBarr}       & SCU           &   & \appIXSFRsizeSCU        & & \appIXenergySCU      & & \appIXexCycTotSCU   & & \appIXexCycActSCU   & (\appIXexCycActSprSCU  ) & & \appIXsyncCycTotSCU   & (\appIXsyncCycTotRelSCU  ) & & \appIXsyncCycActSCU    & (\appIXsyncCycActRelSCU  ) & & \appIXIPCSCU \\
                              &                                 &                                      & TAS           &   & \appIXSFRsizeTAS        & & \appIXenergyTAS      & & \appIXexCycTotTAS   & & \appIXexCycActTAS   & (\appIXexCycActSprTAS  ) & & \appIXsyncCycTotTAS   & (\appIXsyncCycTotRelTAS  ) & & \appIXsyncCycActTAS    & (\appIXsyncCycActRelTAS  ) & & \appIXIPCTAS \\
                              &                                 &                                      & SW            &   & \appIXSFRsizeSW         & & \appIXenergySW       & & \appIXexCycTotSW    & & \appIXexCycActSW    & (\appIXexCycActSprSW   ) & & \appIXsyncCycTotSW    & (\appIXsyncCycTotRelSW   ) & & \appIXsyncCycActSW     & (\appIXsyncCycActRelSW   ) & & \appIXIPCSW  \\
\bottomrule
\end{tabular}
\end{center}
\end{table*}

In relation to the barrier results, the differences between the \gls{SCU} and the \gls{TAS}-based variants are considerably smaller: As can be seen in the right half of \figref{fig:results_synthetic}, for two cores the minimum \gls{SFR} size for 10\% overhead is at most reduced by \edgeSFRenergyMutXratioTAScoreII$\times$ from \edgeSFRenergyMutXTAScoreII\,cycles to \edgeSFRenergyMutXSCUcoreII\,cycles when comparing the energy overhead of the \gls{TAS} and \gls{SCU} variants. The differences in the relative cycle overhead are smaller or even non-existent. The picture changes, however, when considering eight participating cores: While the cycle-related differences remain small, the energy-related gap widens. The smallest \gls{SFR} for 10\% relative energy overhead is reduced by at least \edgeSFRenergyMutXratioSWcoreVIII$\times$ (10-cycle critical section, SW to \gls{SCU}) and up to \edgeSFRenergyMutXratioTAScoreVIII$\times$ (10-cycle critical section, \gls{TAS} to \gls{SCU}). Still, compared to the barrier, the savings achievable with the \gls{SCU} mutex extension are one order of magnitude lower. The reason for this behavior is twofold: First, a mutex is a much simpler \sy primitive than a barrier, and second, a \gls{TAS}-protected variable inherently allows for very efficient implementations. Still, the avoidance of any \gls{TCDM} accesses when using the \gls{SCU} results in consistently lower power and energy for all core counts and critical section lengths.

Counterintuitively, the \gls{TAS}-variant performs both in terms of cycles and energy worse than the straight-forward SW version for all core counts and critical section lengths.
This circumstance can be explained by analyzing the software footprint of each implementation variant: With fully inlined functions for entering and leaving critical sections, leaving always requires the execution of a single instruction for the SW and \gls{SCU} variants and two for the \gls{TAS} variant. For entering, however, only the \gls{SCU} variant guarantees a single instruction for all cores. The naive SW variant requires two instructions per locking attempt; the \gls{TAS} variant can match this count only if the first attempt is successful. For all additional ones, five instructions need to be executed to handle the idle-wait functionality. Conclusively, the \gls{TAS} variant can reduce the number of lock attempts; however, each attempt is more expensive. For cases where re-election takes place after roughly ten cycles, this can thus lead to an overall increase in both cycles and energy used for the primitive that outweigh the energy saved with cores that sleep for very short instances only. Hence, the critical section lengths used in our experiments are simply too short for the \gls{TAS} variant to show a benefit over the SW one; without the \gls{SCU} programmers have to choose the optimal implementation in dependency of $T_{crit}$. Additionally, the usage of \texttt{nop} instructions during the critical section hides a disadvantage of the SW implementation that would show with real applications: The repeated \sy variable polling by all cores that yet have to enter the critical section puts a significant load on both the \gls{TCDM} and the associated interconnect that would slow down the execution of any critical section which contains \gls{TCDM} accesses.

\subsection{DSP Applications}\label{sec:exp_apps}
After exploring the theoretically achievable improvements with dummy code between \sy points, we ran actual \gls{DSP}-centric applications on the multicore cluster, each with the three different implementations of \sy primitives. The applications are, e.g., in turn, applied in real-world use-cases such as \cite{Kartsch19,Palossi19}. Compilation was done in the same way as with the synthetic benchmarks from \secref{sec:synth_bench}; additionally, the combination of the GCC-flags \texttt{-flto} and \texttt{-fno-tree-loop-distribute-patterns} that yields best performance (for each application individually) has been determined and applied. In order to obtain accurate results, each application was run seven times on the \gls{RTL} model, where the two first iterations are used to warm the instruction cache and are not counted towards any results. All cycle-based results are calculated from the averaged outputs of the observation tools over the last five iterations. For calculating power results, the applications were run in the same manner on the post-layout model with signal activity being recorded during a cache-hot iteration. As with the synthetic experiments, all results reflect running the cluster at 350\,MHz.

\begin{figure}[!t]
\centering{
  \includegraphics[width=\columnwidth]{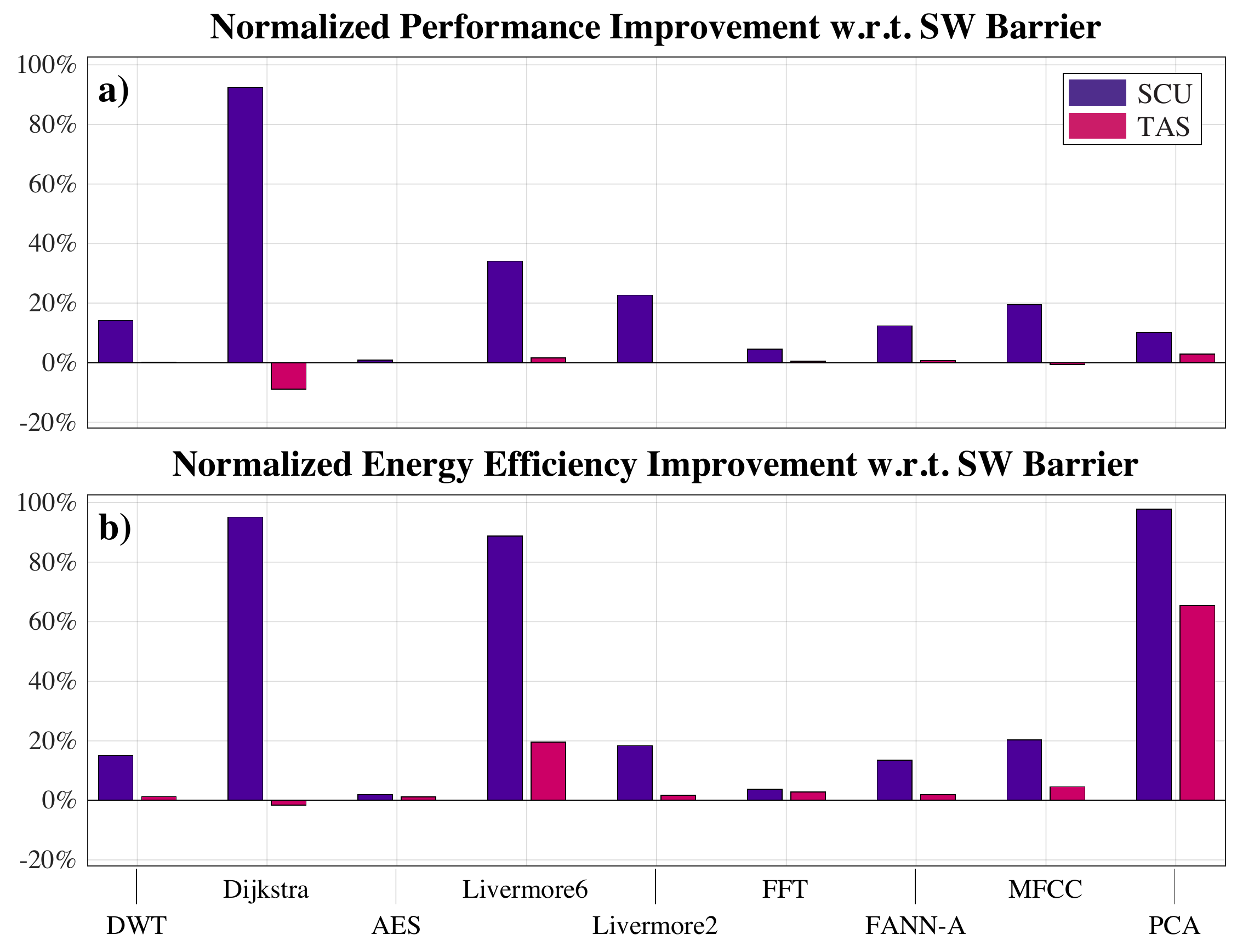}
  \caption{\label{fig:results_norm} Normalized performance (a) and energy improvements (b) for the range of \gls{DSP}-applications and the \gls{SCU} and \gls{TAS} barrier implementations relative to the SW baseline.}
  }
\end{figure}

A short description of each application and its \sy behavior is as follows: \textbf{\appIname:} 512-element 1D Haar real-valued 32-bit fixed point \gls{DWT}; one barrier after the initial variable and pointer setup phase and after each DWT step. \textbf{\appIIname:} Dijkstra's minimum distance algorithm for a graph with 121 nodes; for each node, the minimum distance to node zero is calculated. Two barriers per node that ensure each core is done with its part of the graph before deciding on the minimum distance for each node. \textbf{\appIIIname:} One round of encryption and one round of decryption of 1\,kByte of data using the \gls{AES} in counter mode. Barriers are only used before and after the two phases of the algorithm, as it can be fully vectorized. \textbf{\appIVname:} General linear recurrence equation from the Livermore Loops \cite{Feo88}; the transformed, parallelizable version of the algorithm proposed in \cite{Sampson06} was used with a 128-bit single-precision input vector. A barrier must be passed on each iteration of the outer loop as there are data dependencies between the iterations. \textbf{\appVname:} Excerpt from an incomplete Cholesky-Conjugate gradient descent that processes an 8\,kByte single-precision input vector. The algorithm reduces the part of the vector that is processed in each iteration by a factor of two and, therefore, only requires 12 outer loop iterations after each of which a barrier is required. \textbf{\appVIname:} 512-point complex-valued single-precision radix-8 \gls{FFT} with precomputed twiddle factors. Barriers are only required between each radix-8 butterfly step (two with the input size at hand) and at the end of the algorithm to arrange the output values in the correct order. \textbf{\appVIIname:} Hand gesture recognition from \cite{Wang19}, based on a 32-bit fixed-point fully-connected \gls{FANN} with five layers, 691 neurons, and over 0.4\,MByte of weights. Barriers are required both after processing each layer (outer loop) as well as after each fully-parallel inner loop iteration in which each core calculates a neuron value. The barrier at the inner loop is required to manage the loading of the currently required weight values into the \gls{TCDM} by the \gls{DMA} in the background as the \gls{TCDM} is far too small to fit all values at once. \textbf{\appVIIIname:} Calculation of the \gls{MFC} (inverse \gls{FFT} of the logarithm of the power spectrum) of a 20.000-element 16-bit fixed-point vector. An outer loop runs over frames of four bytes with a barrier after each iteration. For each frame, nine processing steps with a barrier in between each are carried out. With the exception of the forward \gls{FFT} to compute the power spectrum, all processing steps are fully vectorized and free from \sy points. \textbf{\appIXname:} 32-bit fixed-point \gls{PCA} based on Householder rotations on a dataset composed of 23 channels and 256 observations; the algorithm is distributed over five processing steps (data normalization, Householder reduction to bidiagonal form, accumulation of the right-hand transformation, diagonalization, final computation of principal components) with a barrier in between each. Four of the processing steps contain numerous barriers due to data dependencies and short sequential sections for combining intermediate results from preceding parallel sections; the diagonalization part of the algorithm is largely sequential.

The applications were selected with a focus on covering both a wide range of domains and the relevant parameter space: As \tblref{tbl:app_numbers} shows, barrier count, the number of total cycles as well as total energy all range over four orders of magnitude. The range of average \gls{SFR} sizes is roughly lower bound at around 100 cycles (\appIIname), a size for which \figref{fig:results_synthetic}a) and d) show that \sy overheads achieved with the \gls{SCU} barrier are still well below the acceptable margin of 10\%. At the upper end of the spectrum, applications with \gls{SFR} sizes of one thousand cycles and more are as well included (\appIIIname, \appVIname), representing the range of \gls{SFR} sizes where \figref{fig:results_synthetic}a) and d) indicate only small overhead reductions when comparing the \gls{SCU} barrier to the \gls{TAS} and SW baselines.

Both discussed Livermore Loops were mainly chosen for benchmarking to allow for quantitative comparison to literature. We could identify only two cases where systems from literature performed better, however in both cases with at least doubled core count compared to our cluster and only when comparing to the \gls{TAS} or SW type primitives (they resulted in uniformly very similar cycle counts for both Livermore loops): For Livermore6 with data size 256, the 16-core \gls{CMP} from \cite{Sampson06} performs \LivVICycleSampSWbet better; for Livermore2 with a 2048-element vector, the 128-core system of \cite{Sartori10} achieves \LivIICycleSartSWbet lower cycle count. For Livermore2 and all other vector sizes used in \cite{Sampson06,Sartori10}, we achieved performance improvements between \LivIICycleSartTASmin and \LivIICycleSartTASmax with the \gls{TAS} and SW barrier variants and between \LivIICycleSartSCUmin and \LivIICycleSampSCUmax with the \gls{SCU} barrier. For Livermore6, the improvements in comparison to \cite{Sampson06} range between \LivVICycleSampSWmin and \LivVICycleSampTASmax with the \gls{TAS} and SW barriers and between \LivVICycleSampSCUmin and \LivVICycleSampSCUmax when using \gls{SCU} barriers.

In relation to the 7-core system used in \cite{Xiao12}, performance for Livermore2 was improved by over \LivIICycleXiaoSW with \gls{TAS} and SW barriers and by over \LivIICycleXiaoSCU with the \gls{SCU} barrier. For Livermore6, we observe an improvement of \LivVICycleXiaoSW with the \gls{TAS} and SW type primitive and \LivVICycleXiaoSCU when using the \gls{SCU}. For both benchmarks, \cite{Xiao12} uses a 1024-element vector.

\begin{figure*}[!t]
\centering{
  \includegraphics[width=1\textwidth]{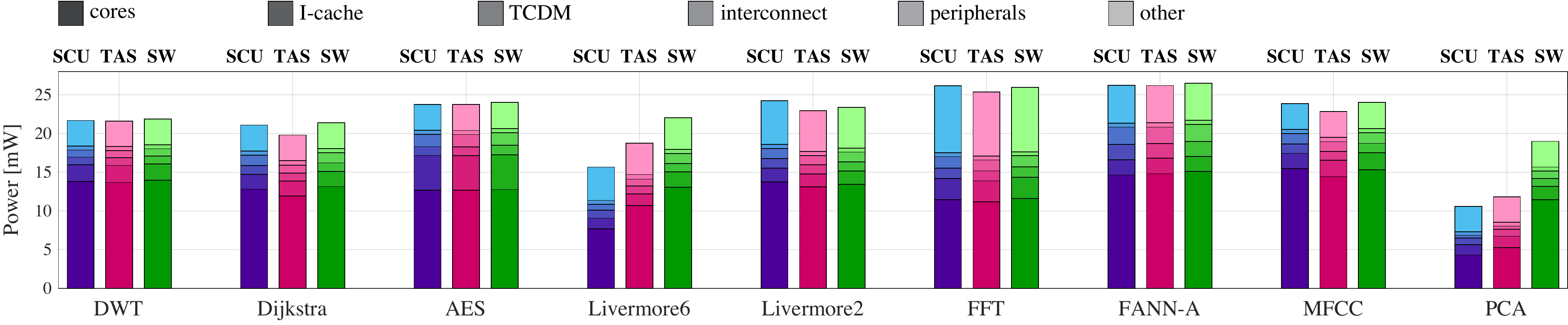}
  \caption{\label{fig:apps_power} Total cluster power and breakdown into the main contributors.}
  }
\end{figure*}

\subsubsection*{Calculation of Synchronization Overhead}
As the main goal of this work is to boost energy efficiency by drastically reducing the \sy-related overhead, we provide for each application and \sy primitive implementation both the number of total and active cycles that cores use to execute \sy primitives.
The cycle counts have been determined with a profiling script that parses the trace files of each core and application. For \gls{SCU}-type \sy primitives, the detection is done by searching for the \texttt{elw} instruction with matching physical address. Any preceding instructions that are used to calculate the address are as well counted towards the \sy cycles. In the case of the \gls{TAS} and SW variants, two detection methods have to be used: By analyzing the disassembly of each application, the address range(s) of \sy functions are extracted. If this step succeeds, the traces can successively simply be scanned for time periods where a core executes instructions within a relevant address range. For many applications, however, this method fails due to the fact that the compiler inlines \sy functions. Consequently, the inlined functions must be detected by matching the disassembly against patterns that unambiguously identify \sy primitives. This method requires much more careful analysis as the functions can be spread across multiple non-contiguous address ranges with linking jump or branch instructions. Furthermore, multiple entry- and exit points to and from the primitives may exist. The output of the described analysis methods is, in any case, a list of \sy periods where each entry contains a begin and end cycle number. Combining these timestamps with the benchmarking intervals allows us to calculate both the total and active number of \sy cycles for each core and benchmark iteration, the average of which is shown in \tblref{tbl:app_numbers}.

It is important to note that the \textit{total} number of \sy cycles naturally includes core wait periods that are mostly caused by workload imbalance. Therefore, the number of \textit{active} \sy cycles is a much better measure for the actual \sy overhead and substantially lower than the former count for the idle-wait featuring \gls{SCU} and \gls{TAS} variants and considering applications that exhibit significant workload imbalance (\appIVname, \appIXname).

\subsubsection*{Discussion of Results}
\tblref{tbl:app_numbers} lists the most important properties of each application alongside the cycle-based and energy results. In order to provide full insight into the components contributing to energy, \figref{fig:apps_power} shows both total power and the corresponding breakdown into the shares associated with the main cluster components. Finally, \figref{fig:results_norm} highlights the normalized improvements in terms of cycles and energy that we were able to achieve with each application when employing the \gls{SCU}- and \gls{TAS}-based \sy primitives in relation to the SW baseline. Over the range of the benchmarked applications, the \gls{SCU} achieves relative performance improvements between \appRelCycleSCUmin and \appRelCycleSCUmax, with an average of \appRelCycleSCUavg. While the lower and upper bounds for relative energy improvement are very similar, amounting to \appRelEnergySCUmin and \appRelEnergySCUmax, respectively, power reductions with the \gls{SCU} favorably result in a greater average improvement of \appRelEnergySCUavg.

When relating the average \gls{SFR} size from \tblref{tbl:app_numbers} with the normalized improvements from \figref{fig:results_norm}, one can see that the \gls{SFR} size is a strong indicator whether the type of \sy implementation influences overall performance and energy or not. Consequently, the biggest improvements are achieved with applications that exhibit \glspl{SFR} of few hundreds of cycles (\appIIname, \appIVname, \appIXname) and the lowest with \glspl{SFR} sizes of thousand or several thousands of cycles (\appIIIname, \appVIname). It can be noticed that the results of the synthetic benchmarks in \secref{sec:synth_bench}, shown in \figref{fig:results_synthetic}, can provide a rough estimate of the achievable savings: The relative overhead of \textit{active} \sy cycles (averaged over all cores) for each application and barrier variant is marked in \figref{fig:results_synthetic}a. Instead of the number of total \sy cycles, the amount of \textit{active} \sy cycles is used since, in the synthetic benchmarks, all cores arrive at almost the same time at a barrier while in real applications, core-to-core workload imbalances cause a much higher variation of the arrival instances. For the \gls{SCU} barrier, the overhead predicted by the synthetic benchmarks closely matches the actual application-related one for all applications; for the \gls{TAS} and SW barriers, however, the synthetic experiments mostly predict overheads that are significantly too high. This can be explained with the already mentioned workload imbalances; the spread-out arrival instances also reduce core-concurrent access to the \gls{TAS}-variable that protects the barrier status from hazardous modification. As a consequence, fewer cycles are wasted due to contention while accessing the said variable. The fully-parallel access to the \gls{SCU} barrier extension, on the other hand, causes the barrier durations to be completely independent of the distribution of the arrival instances, leading to a greatly improved (cycle) overhead predictability.

An important observation is the fact, that the \gls{SCU}, -- for most applications -- does \textit{not} reduce power but either almost does not affect it at all (see \appIname, \appIIIname, \appVIIname) or even slightly increases power compared to the \gls{TAS} primitive variant, which also features idle-waiting (see \appIIname, \appVname, \appVIname, \appVIIIname). As \figref{fig:apps_power} shows, the increase in the latter case is due to higher power consumption in the cores, which is a consequence of the reduction of \sy cycles and the relative higher share of (usually) energy-intensive processing cycles.
There are, however, two exceptions to this behavior, \appVname\xspace and \appIXname, where total power is reduced by \appIVpowRedTAS and \appIXpowRedTAS when using the \gls{TAS} barrier, or, respectively, \appIVpowRedSCU and \appIXpowRedSCU with the \gls{SCU} variant. In these cases, application-inherent workload imbalances indicated through the standard deviation of active execution cycles (over cores) in \tblref{tbl:app_numbers}, result in large differences for the times that individual cores wait at barriers. Avoiding active spinning on \sy variables during the resulting prolonged wait periods with both the \gls{TAS} and \gls{SCU} barriers reduces the power of the involved components (cores, interconnect, \gls{TCDM}) and -- since they consume the lion share of overall power -- also of the whole cluster very significantly. This circumstance also shows when comparing the normalized cycle and energy efficiency improvements in \figref{fig:results_norm}, where the gains in energy efficiency are very similar to those for performance except for the two applications in question; in those cases, the discussed power reduction results in much greater improvements for energy efficiency.

This work focuses on the optimization of the core-to-core communication and \sy on ultra-low-power clusters of processors in the \gls{IoT} domain, leveraging parallelism to improve energy-efficiency of computations rather than performance only. In different contexts, such as high-end devices, having power/performance scalable systems, able to scale-up to 100s or 1000s of cores, is a desirable feature. However, state-of-the-art parallel computing systems, such as GP-GPUs, feature a clear trade-off between performance and efficiency, both from the point of view of the parallelism available in embedded applications as well as from a physical implementation perspective (as discussed in \secref{sec:background}). In the context of PULP-based systems, where energy-efficiency cannot be traded off against performance, scalability is still an open problem, and we plan to explore this scenario as future work.

\section{Conclusion}\label{sec:conclusion}
We proposed a light-weight hardware-supported \sy concept for embedded \glspl{CMP} that aggressively reduces \sy overhead, both in terms of execution time and energy, the latter being a crucial metric for most embedded systems. In addition to showing energy cost reductions for \sy primitives of up to \energyBarrRatioTAScoreVIII$\times$ and resulting minimum \gls{SFR} sizes of as little as tens of cycles, we demonstrated the importance of energy-efficient \sy on a range of typical applications that covers four orders of magnitude of execution time and \gls{SFR} size. The proposed solution improves both performance and energy efficiency in all cases and has a beneficial impact of up to \appRelCycleSCUmax for performance and \appRelEnergySCUmax for energy efficiency for applications with \gls{SFR} sizes of around one hundred cycles. In the future, we plan to explore hierarchical architectures composed of multiple tightly-coupled clusters, with the target of scaling up the performance of PULP systems with no compromises on energy-efficiency.

\ifCLASSOPTIONcaptionsoff
  \newpage
\fi



%
\bibliographystyle{IEEEtran}
\bibliography{SCU_TPDS}

%
%

%


%
%
\begin{IEEEbiography}[{\includegraphics[width=1in,height=1.25in,clip,keepaspectratio]{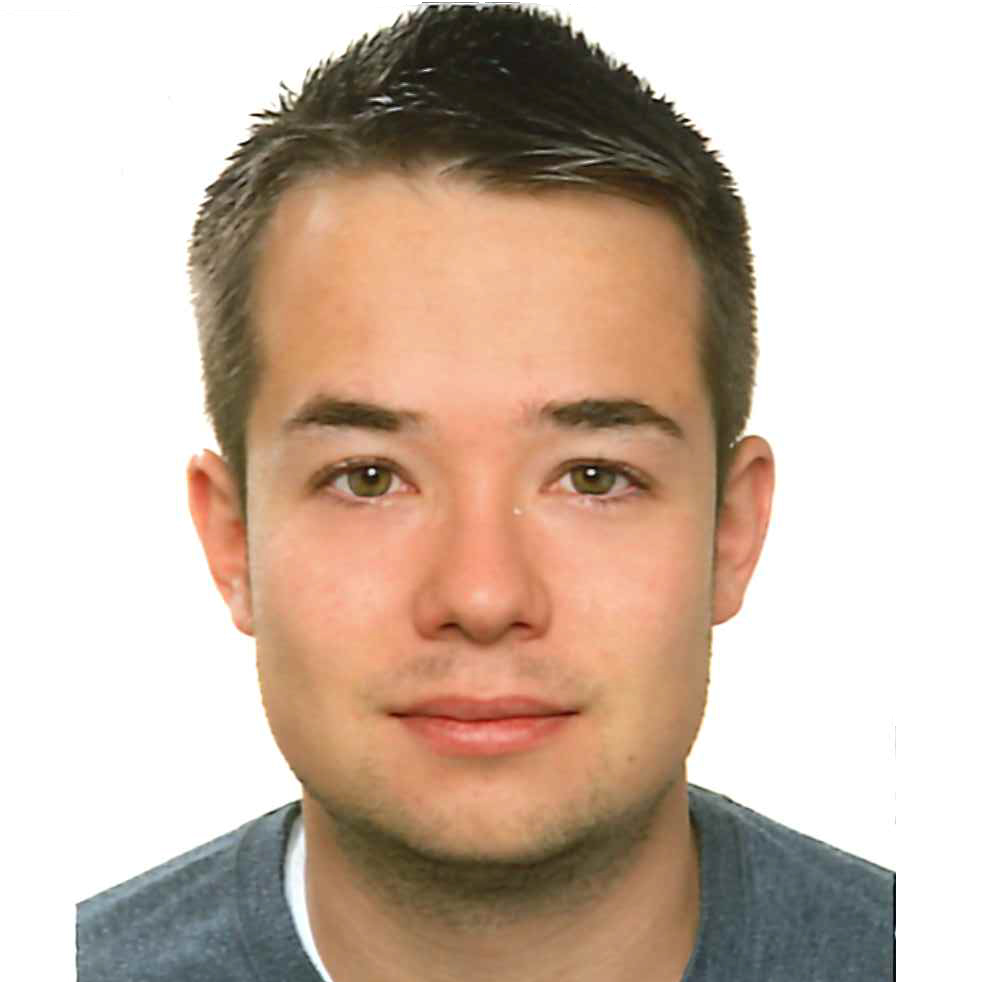}}]{Florian Glaser}
 received the M.Sc. degree in electrical engineering from ETH Zurich, Switzerland, in 2015, where he is currently pursuing the Ph.D. degree at the Integrated Systems Laboratory. His current research interests include low-power integrated circuits with a special focus on energy-efficient synchronization of multicore clusters and mixed-signal systems-on-chip for miniaturized biomedical instrumentation.
\end{IEEEbiography}
\begin{IEEEbiography}[{\includegraphics[width=1in,height=1.25in,clip,keepaspectratio]{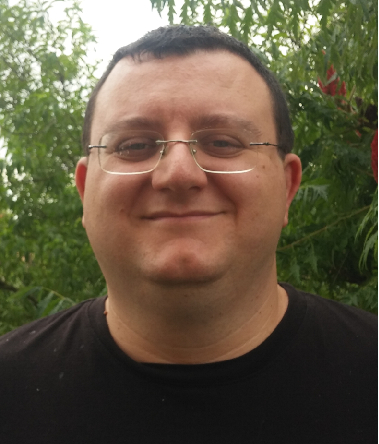}}]{Giuseppe Tagliavini}
received the Ph.D. degree in electronic engineering from the University of Bologna, Bologna, Italy, in 2017.
He is currently a Post-Doctoral Researcher with the Department of Electrical, Electronic, and Information Engineering, University of Bologna. He has coauthored over 20 papers in international conferences and journals. His research interests include parallel programming models for embedded systems, run-time optimization for multicore and many-core accelerators, and design of software stacks for emerging computing architectures.
\end{IEEEbiography}
\begin{IEEEbiography}[{\includegraphics[width=1in,height=1.25in,clip,keepaspectratio]{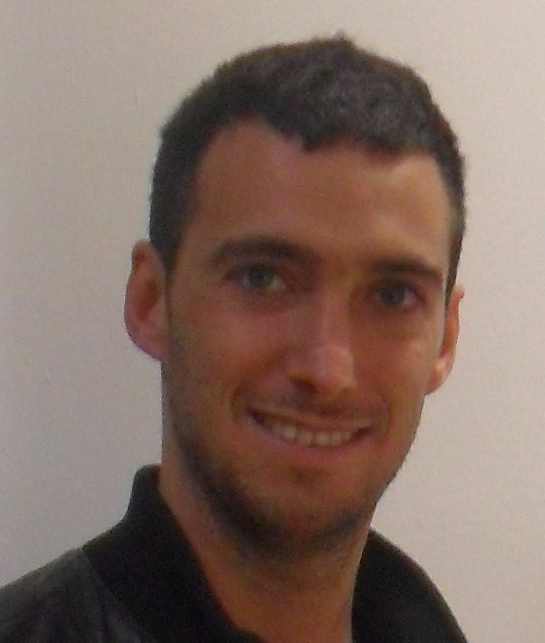}}]{Davide Rossi}
received the Ph.D. degree from the University of Bologna, Bologna, Italy, in 2012.
He has been a Post-Doctoral Researcher with the Department of Electrical, Electronic and Information Engineering “Guglielmo Marconi,” University of Bologna, since 2015, where he is currently an Assistant Professor. His research interests focus on energy-efficient digital architectures. In this field, he has published more than 80 papers in international peer-reviewed conferences and journals.
\end{IEEEbiography}
\begin{IEEEbiography}[{\includegraphics[width=1in,height=1.25in,clip,keepaspectratio]{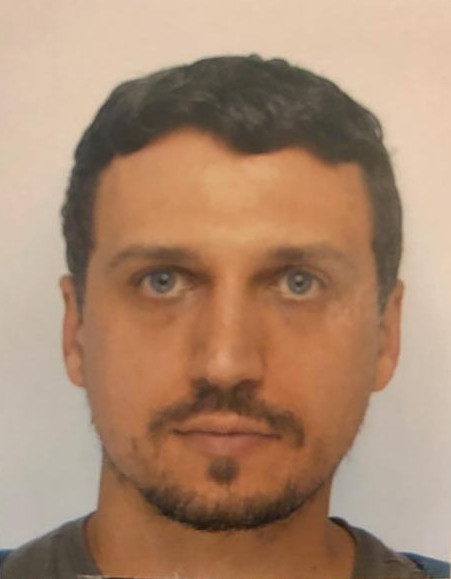}}]{Germain Haugou}
received the Engineering Degree in telecommunication from the University of Grenoble, in 2004. He was with ST Microelectronics as a Research Engineer, for ten years. He is currently with ETH Zurich, Switzerland, as a Research Assistant. His research interests include virtual platforms, run-time systems, compilers, and programming models for many-core embedded architectures.
\end{IEEEbiography}
\begin{IEEEbiography}[{\includegraphics[width=1in,height=1.25in,clip,keepaspectratio]{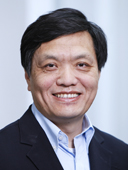}}]{Qiuting Huang}
	received the Ph.D. degree in applied sciences from the Katholieke Universiteit Leuven, Leuven, Belgium, in 1987.
	
	Between 1987 and 1992, he was a lecturer at the University of East Anglia, Norwich, UK. Since January 1993, he has been with the Integrated Systems Laboratory, ETH Zurich, Switzerland, where he is Professor of Electronics. In 2007, he was also appointed as a part-time Cheung Kong Seminar Professor by the Chinese Ministry of Education and the Cheung Kong Foundation and has been affiliated with the South East University, Nanjing, China. His research interests span RF, analog, mixed analog-digital as well as digital application specific integrated circuits and systems, with an emphasis on wireless communications and biomedical applications in recent years.
	
	Dr. Huang currently serves as vice chair of the steering committee, as well as a sub committee chair of the technical program committee of the European Solid-State Circuits Conference (ESSCIRC). He also served on the technical program and executive committees of the International Solid-State Circuits Conference (ISSCC) between 2000 and 2010.
\end{IEEEbiography}
\begin{IEEEbiography}[{\includegraphics[width=1in,height=1.25in,clip,keepaspectratio]{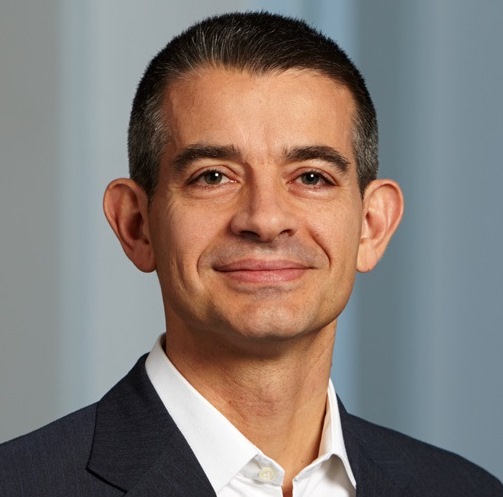}}]{Luca Benini}
holds the chair of the Digital Circuits and Systems Group at the Integrated Systems Laboratory, ETH Zurich and is Full Professor at the Universita di Bologna.
He has been visiting professor at Stanford University, IMEC, and EPFL. He served as chief architect in STMicroelectronics France.

Dr. Benini's research interests are in energy-efficient parallel computing systems, smart sensing micro-systems and machine learning hardware.
He has published more than 1000 peer-reviewed papers and five books.

He is a Fellow of the IEEE, of the ACM and a member of the Academia Europaea.
He is the recipient of the 2016 IEEE CAS Mac Van Valkenburg award and of the 2019 IEEE TCAD Donald O. Pederson Best Paper Award. 
\end{IEEEbiography}
%
%







\end{document}